\documentclass[aps,prd,10pt,twocolumn,nofootinbib,groupedaddress,superscriptaddress,floatfix]{revtex4-1}

\usepackage[colorlinks=true,urlcolor=blue,linkcolor=,citecolor=blue]{hyperref}
\usepackage{amsmath,amssymb,amstext,amsbsy,amsfonts,amsthm,graphicx,microtype,booktabs}
\usepackage[capitalize]{cleveref}

\newcommand{\abs}[1]{{\left \vert #1 \right \vert}}
\newcommand{\ud}{\textrm{d}}
\newcommand{\Mpl}{M_\mathrm{pl}}

\begin{document}

\title{Constraining axion inflation with gravitational waves from preheating}

\author{Peter Adshead}
\email{adshead@illinois.edu}
\affiliation{Department of Physics, University of Illinois at Urbana-Champaign, Urbana, Illinois 61801, USA}

\author{John T. Giblin, Jr.}
\email{giblinj@kenyon.edu}
\affiliation{Department of Physics, Kenyon College, Gambier, Ohio 43022, USA}
\affiliation{CERCA/ISO, Department of Physics, Case Western Reserve University, Cleveland, Ohio 44106, USA}

\author{Mauro Pieroni}
\email{mauro.pieroni@uam.es}
\affiliation{Instituto de F\'{\i}sica Te\'orica UAM/CSIC, Calle Nicol\'as Cabrera 13-15, Cantoblanco E-28049 Madrid, Spain}
\affiliation{Departamento de F\'{\i}sica Te\'orica, Universidad Aut\'onoma de Madrid (UAM) Campus de Cantoblanco, 28049 Madrid, Spain}
\affiliation{Theoretical Physics, Blackett Laboratory, Imperial College, London, SW7 2AZ, United Kingdom}

\author{Zachary J. Weiner}
\email{zweiner2@illinois.edu}
\affiliation{Department of Physics, University of Illinois at Urbana-Champaign, Urbana, Illinois 61801, USA}

\begin{abstract}
    We study gravitational wave production from gauge preheating in a variety of inflationary models, detailing its dependence on both the energy scale and the shape of the potential.
    We show that preheating into Abelian gauge fields generically leads to a large gravitational wave background that contributes significantly to the effective number of relativistic degrees of freedom in the early universe, $N_\mathrm{eff}$.
    We demonstrate that the efficiency of gravitational wave production is correlated with the tensor-to-scalar ratio, $r$.
    In particular, we show that efficient gauge preheating in models whose tensor-to-scalar ratio would be detected by next-generation cosmic microwave background experiments ($r \gtrsim 10^{-3}$) will be either detected through its contribution to $N_\mathrm{eff}$ or ruled out.
    Furthermore, we show that bounds on $N_\mathrm{eff}$ provide the most sensitive probe of the possible axial coupling of the inflaton to gauge fields regardless of the potential.
\end{abstract}

\maketitle


\section{Introduction}\label{sec:Introduction}

While the qualitative predictions of inflation are well motivated and well understood~\cite{Guth:1980zm, Linde:1981mu, Albrecht:1982wi, Linde:1983gd, Akrami:2018odb}, there is as yet no unique, complete model which connects inflation to the standard model (SM) of particle physics.
A crucial component of such a model is the subsequent reheating process~\cite{Traschen:1990sw, Shtanov:1994ce,Kofman:1994rk,Kofman:1997yn}, which must realize a phase transition from the cold postinflationary state to the hot Big Bang.
In this phase, the inflaton decays into other species to repopulate the Universe and begin the radiation-dominated era.

During an initial stage of \textit{preheating}, the coherent oscillation of the inflaton field induces explosive production of bosons via parametric or tachyonic resonance (see Refs.~\cite{Amin:2014eta,Allahverdi:2010xz} for reviews).
The rapid production of inhomogeneities during this phase generically sources a significant gravitational wave background~\cite{Khlebnikov:1997di, Easther:2006gt, Easther:2006vd, Easther:2007vj, GarciaBellido:2007dg, Dufaux:2007pt, Dufaux:2010cf, Bethke:2013aba, Figueroa:2013vif, Bethke:2013vca, Figueroa:2016ojl, Figueroa:2017vfa}.
Furthermore, preheating can have important consequences for observable predictions of inflationary models in the cosmic microwave background (CMB): the evolution of the equation of state after inflation affects the mapping of fluctuations' present-day length scales to the times those scales first left the horizon during inflation.
In particular, if preheating is sufficiently efficient then the onset of radiation domination occurs nearly instantaneously after inflation ends.

Most of the early work on preheating focused on models which couple the inflaton to another canonical scalar field~\cite{GarciaBellido:1997wm,Khlebnikov:1997di,Greene:1997ge,Parry:1998pn,Bassett:1998wg,GarciaBellido:1998gm,Easther:1999ws,Liddle:1999hq,Finelli:2001db,Bassett:2005xm,Podolsky:2005bw}, with recent studies exploring nonminimally coupled scalar fields~\cite{Child:2013ria,Fu:2017ero,Nguyen:2019kbm,Crespo:2019src,Fu:2019qqe}.
Alternatively, the inflaton could couple directly to gauge fields~\cite{Deskins:2013dwa,Adshead:2015pva,Adshead:2016iae,McDonough:2016xvu,Lozanov:2016pac,Giblin:2017wlo,Adshead:2017xll,Cuissa:2018oiw,Adshead:2018doq}.
One such coupling of particular theoretical and phenomenological interest is that of a (pseudoscalar) inflaton $\phi$ coupled to the Chern-Simons density $F \tilde{F}$ of a gauge field~\cite{Turner:1987bw, Garretson:1992vt, Anber:2006xt}.
From the theoretical point of view, the (approximate) shift symmetry of a pseudoscalar inflaton (axion) inherently protects its potential from large corrections, ensuring the flatness required for a successful inflationary phase.
The phenomenology of these models is extremely rich, with possible observable signatures including the production of sizable non-Gaussianities~\cite{Barnaby:2010vf, Barnaby:2011qe, Barnaby:2011vw, Anber:2012du, Linde:2012bt}, observable gravitational waves~\cite{Cook:2011hg, Barnaby:2011qe, Anber:2012du, Domcke:2016bkh, Bartolo:2016ami, Jimenez:2017cdr}, primordial black holes~\cite{Linde:2012bt, Bugaev:2013fya, Garcia-Bellido:2016dkw, McDonough:2016xvu, Domcke:2017fix, Garcia-Bellido:2017aan, Cheng:2018yyr}, $\mu$-distortions~\cite{Meerburg:2012id, Domcke:2016bkh}, primordial magnetic fields~\cite{Anber:2006xt, Durrer:2010mq, Caprini:2014mja, Fujita:2015iga, Green:2015fss, Patel:2019isj}, and the generation of the baryon asymmetry~\cite{Anber:2015yca, Cado:2016kdp, Jimenez:2017cdr, Domcke:2018eki, Domcke:2019mnd}.

Preheating into gauge fields via a Chern-Simons coupling  was first studied within the context of chaotic inflation in Ref.~\cite{Adshead:2015pva}, and the model's viability for magnetogenesis was explored in Ref.~\cite{Adshead:2016iae}.
Subsequently, we extended this work in Ref.~\cite{Adshead:2018doq} to compute the gravitational wave spectrum produced by the dynamics of gauge preheating.
The surprising result was that the (over) production of gravitational radiation provides the strongest probe of (or constraints on) the coupling scale between the axion and gauge fields.
As detailed in Ref.~\cite{Adshead:2016iae} (and \cref{sec:gw-neff} below), next-generation CMB experiments will probe the radiation content of the Universe to a precision sufficient to rule out most of the interesting region of parameter space in these models.
Here and in Ref.~\cite{Adshead:2019igv} we use this result to put important constraints on the axial coupling between the inflaton and gauge fields in a variety of well-motivated inflationary models.

While the chaotic inflation scenario considered in Ref.~\cite{Adshead:2018doq} models the inflaton's coherent oscillations about the minimum of its potential to leading order, this model is disfavored at the 95\% confidence level, primarily by constraints on the tensor-to-scalar ratio~\cite{Akrami:2018odb}.
To more completely understand the role gauge preheating could play in constraining axion inflation, we investigate the dependence of preheating dynamics and gravitational wave production on the details of the inflationary potential.
While the efficiency of preheating and the production of gravitational waves is qualitatively generic, both the energy scale of inflation and the shape of the potential significantly alter the quantitative results as we demonstrate below.

This paper is organized as follows.
In \cref{sec:Model} we review models in which an axion or pseudoscalar inflaton is coupled to an Abelian gauge field and introduce the constraints on stochastic backgrounds of gravitational waves from the CMB.
After describing the numerical prescription employed for simulations in \cref{sec:numerical-methods}, we outline some analytic estimates in \cref{sec:analytic-estimates}.
We investigate the effect of the energy scale of inflation and the shape of the inflationary potential on preheating and gravitational wave production in \cref{sec:dependence-inflationary-scale,sec:dependence-potential-shape}, respectively.
In \cref{sec:Conclusions} we draw our conclusions.
In \cref{App:eomsandlinalaysis} we detail the equations of motion of the system and summarize a linear analysis thereof.
\Cref{sec:initial-conditions} details our procedure for setting accurate initial conditions.
Finally, in \cref{sec:post-preheating-dynamics} we verify that the Universe remains radiation dominated after preheating to establish the robustness of the gravitational-wave transfer function (for the models we consider).


\section{Gauge fields during and after axion inflation}\label{sec:Model}

In this section we introduce the models of interest and briefly review the dynamics of axially coupled gauge fields during and after axion inflation, relegating many details to \cref{App:eomsandlinalaysis}.


\subsection{Models of inflation}\label{sec:modelsub}

We consider the action for a pseudoscalar inflaton $\phi$ minimally coupled to gravity and to an Abelian gauge field $A_{\mu}$,
\begin{align}
    \begin{split}
    \label{eqn:action}
        S
        &= \int \ud^4 x \sqrt{-g} \Bigg[
            \frac{\Mpl^2}{2} R - \frac{1}{2} \partial_\mu \phi \partial^\mu \phi - V(\phi) \\
        &\hphantom{={} \int \ud^4 x \sqrt{-g} \Bigg[ }
            - \frac{1}{4} F_{\mu\nu} F^{\mu\nu} - \frac{\alpha}{4 f} \phi F_{\mu\nu}\tilde{F}^{\mu\nu}
            \Bigg],
    \end{split}
\end{align}
where $f$ is a mass scale (the axion decay constant), $\alpha$ is a dimensionless coupling constant, and $F_{\mu\nu} \equiv \partial_\mu A_{\nu} - \partial_\nu A_{\mu } $ is the field strength tensor of a U(1) gauge theory.\footnote{The extension to different Abelian gauge groups [such as $\mathcal{N}$ copies of a U(1) gauge group] is straightforward.
Conversely, the extension to non-Abelian gauge groups is nontrivial.
See, for example, Refs.~\cite{Maleknejad:2011jw, Adshead:2012kp, Adshead:2012qe, Maleknejad:2012fw, Maleknejad:2016qjz, Adshead:2016omu, Caldwell:2017chz, Domcke:2018rvv, Domcke:2019lxq}.}
We do not identify the gauge field as, e.g., that of standard model hypercharge, nor do we include any charged fields in our model.
The dual tensor is defined by
\begin{align}
    \tilde{F}^{\mu \nu}
    &= \frac{1}{2} \epsilon^{\mu \nu \alpha \beta} F_{\alpha \beta}.
\end{align}
where $\epsilon^{\mu\nu\rho\sigma}$ is the Levi-Civita symbol with convention $\epsilon^{0123} = 1 / \sqrt{-g}$.
We set $c = \hbar = k_B = 1$ and denote by $\Mpl = 1 / \sqrt{8 \pi G_N}$ the reduced Planck mass.
The background spacetime is the mostly plus, conformal Friedmann-Lema\^{\i}tre-Robertson-Walker (FLRW) metric,
\begin{align}
    \label{eqn:flrw-metric}
    \ud s^2 = a{(\tau)}^2 \left( - \ud \tau^2 + \ud {\mathbf{x}}^2 \right),
\end{align}
where the conformal time coordinate $\tau$ is related to the cosmic time coordinate via $\ud \tau = \ud t / a$.
Throughout, primes denote derivatives with respect to conformal time, e.g., $a' \equiv \partial a / \partial \tau$, and dots denote derivatives with respect to cosmic time, e.g., $\dot{a} \equiv \partial a / \partial t$.
We use repeated lower indices to indicate contractions with the Kronecker delta.

In this work, we are interested in the effects of varying the shape of the potential $V(\phi)$ on the efficiency of preheating and the subsequent gravitational wave production.
For this purpose we consider various classes of inflationary potentials (the most representative among the ones typically considered by Planck~\cite{Akrami:2018odb}), reporting our parameter choice(s) and the corresponding scalar tilt $n_s$ and tensor-to-scalar ratio $r$ (evaluated at a pivot scale which exited the horizon 60 $e$-folds before the end of inflation).
It is worth stressing that in the context of standard inflationary model building most of these models assume the inflaton to be a \textit{scalar} field.
Conversely, for the models described by \cref{eqn:action} the inflaton is \emph{pseudoscalar}.
\unskip\footnote{In the context of supergravity this can be achieved by, e.g., using a K\"{a}hler potential which is shift symmetric under the imaginary part of a chiral superfield~\cite{Kawasaki:2000yn}.
In order to generate a potential for the inflaton the shift symmetry must be broken, which can be achieved, for example, by coupling the inflaton to a 3-form~\cite{Dvali:2005an, Kaloper:2008fb, Kaloper:2011jz}.}

Specifically, we explore the following classes of models:
\begin{enumerate}
\item \textit{Chaoticlike monomial models}, which have the potential
	\begin{align}
    	\label{eq:chaotic-class}
	    V(\phi) = m^{4-n} \vert \phi \vert^n.
	\end{align}
    Notice that for $n = 2$ this simply reduces to chaotic inflation~\cite{Linde:1983gd}, the case we consider here.
    For $n = 2$ the Planck normalization of the scalar power spectrum~\cite{Akrami:2018odb} sets $m^2 = 1.9 \times 10^{-11} \, \Mpl^2$, with $n_s = 0.966$ and $r = 0.13$.
    Another set of models, \textit{natural inflation}, have potentials of the form $V(\phi) = V_0 \left( 1 + \cos\left( \phi/v \right) \right)$~\cite{Freese:1990rb}.
    In this case, with $v = \sqrt{8 \pi} \Mpl$ the Planck normalization sets $V_0 = 5.9 \times 10^{-10} \, \Mpl^4$, for which $n_s = 0.952$ and $r = 0.033$.
    However, we find that the results in this case are virtually identical to those for chaotic inflation, so we omit them below.

\item \textit{Starobinsky-like models}, which have potentials given by~\cite{Starobinsky:1980te}\footnote{For other models of this class see also Refs.~\cite{Dvali:1998pa, Burgess:2001vr, Cicoli:2008gp}.}
	\begin{align}
    	\label{eq:starobinsky_class}
	    V(\phi) = V_0 \left( 1 - \exp\left( \frac{\vert \phi \vert}{v} \right) \right)^2.
	\end{align}
    For $v = 10\, \Mpl / 3$ the Planck normalization sets $V_0 = 6.2 \times 10^{-10} \, \Mpl^4$, in which case $n_s = 0.969$ and $r = 0.016$.
    The class of \textit{$\alpha$-attractor models}, motivated by supergravity~\cite{Kallosh:2013yoa,Ferrara:2013rsa} (typically divided into E-models and T-models) have exponentially flat potentials whose steepness is controlled by the parameter $\alpha$, related to the curvature of the K\"{a}hler manifold.
    Since by varying this parameter it is possible to interpolate between chaoticlike models and Starobinsky-like models, we omit the results for this class.

\item  \textit{Monodromy inflation}, corresponding to the potential
    \begin{align}
        V(\phi) = \mu^3 \left( \sqrt{\phi^2 + \phi_c^2} - \phi_c \right).
    \end{align}
    This is part of a broad class of string-theory--motivated models in which large (super-Planckian) field displacement is obtained by wrapping the inflaton trajectory into a series of (sub-Planckian) fundamental circuits~\cite{Silverstein:2008sg, McAllister:2008hb, Flauger:2009ab}.
    Note that deep in the inflationary phase ($\phi \gg \phi_c$) the potential is well approximated by a linear potential ($n=1$) of the class in~\cref{eq:chaotic-class}.
    For $\phi_c = \Mpl / 10$ the Planck normalization sets $\mu = 6.0 \times 10^{-4} \, \Mpl$, for which $n_s = 0.975$ and $r = 0.067$.

\item \textit{Hilltoplike models}, with potentials given by~\cite{Boubekeur:2005zm}
    \begin{align}
	    V(\phi) = V_0 \left( 1 - \left( \frac{\vert \phi \vert}{v} \right)^p \right)^2,
	\end{align}
	where $p \geq 2$.
    We consider a variety of $p$ and $v$ and list the corresponding model fits and predictions in \cref{tab:hilltop-model-fits}.

    \begin{table}[ht]
        \centering
        \caption{
            The combinations of $p$ and $v$ we consider for the hilltop model, the corresponding value of $V_0$ from the Planck normalization, and the corresponding predictions for the scalar tilt and tensor-to-scalar ratio.
            }\label{tab:hilltop-model-fits}
        \begin{tabular*}{\columnwidth}[t]{c @{\extracolsep{\fill}} cccc}
            \toprule
            $p$ &   $v / \Mpl$ &    $V_0 / \Mpl^4$ &        $n_s$ &     $r$ \\
            \midrule
            3 &     $1$ &           $6.5 \times 10^{-16}$ & $0.932$ &  $2.0 \times 10^{-8}$ \\
            4 &     $1$ &           $2.1 \times 10^{-14}$ & $0.949$ &  $6.4 \times 10^{-7}$ \\
            4 &     $2$ &           $3.2 \times 10^{-13}$ & $0.949$ &  $9.8 \times 10^{-6}$ \\
            4 &     $4$ &           $4.7 \times 10^{-12}$ & $0.951$ &  $1.4 \times 10^{-4}$ \\
            4 &     $8$ &           $5.7 \times 10^{-11}$ & $0.955$ &  $1.7 \times 10^{-3}$ \\
            \bottomrule
        \end{tabular*}
    \end{table}

\item \textit{D-brane models}, with potentials of the form~\cite{Dvali:2001fw, Burgess:2001fx, GarciaBellido:2001ky, Kachru:2003sx}
	\begin{align}
	    V(\phi) = V_0 \left( 1 - \left( \frac{v}{\vert \phi \vert} \right)^p \right)^2.
	\end{align}
    Fixing $p = 2$ and $v = \Mpl/2$, the Planck normalization sets $V_0 = 7.5 \times 10^{-11} \, \Mpl^4$, in which case $n_s = 0.975$ and $r = 2.2 \times 10^{-3}$.
 \end{enumerate}


\subsection{Gauge fields on the axion background}

The equations of motion that result from the variation of \cref{eqn:action} are difficult to solve analytically due to the nonlinear interactions of the axion and gauge field.
Furthermore, the homogeneous axion background sources a tachyonic instability in the gauge field which quickly renders a linear analysis invalid.
In this subsection, we briefly outline the well-known linear treatment of the gauge-field fluctuations in the inflationary phase and during early stages of resonance.
Full details may be found in \cref{App:eomsandlinalaysis}.

In a homogeneous axion background, the helical polarizations of the gauge field obey~\cite{Adshead:2015pva}
\begin{align}
    \label{eqn:linear-eom-gauge-polarization}
    {A_\pm}''(\mathbf{k}) + k \left( k \mp 2 \mathcal{H} \xi \right) A_\pm(\mathbf{k}) = 0,
\end{align}
where we define the instability parameter
\begin{align}
    \xi \equiv \frac{\alpha}{2 f} \frac{\phi'}{ \mathcal{H} }
\end{align}
and the conformal Hubble parameter, $\mathcal{H} = a' / a$.
The interaction between the gauge field and the homogeneous, rolling axion leads to the exponential growth of one of the polarizations of the gauge field, understood as an imaginary (``tachyonic'') effective frequency $\omega \sim \sqrt{k(k - 2 \mathcal{H} \xi)}$ (assuming $\xi > 0$) for modes $k < 2\mathcal{H}\xi$.
The polarizations are amplified (relative to the conformally invariant radiation solution $A_{\mathrm{rad},\pm}$) by a factor of
\begin{align}
    \left\vert \frac{A_{\pm}}{A_{\mathrm{rad}, \pm}} \right\vert
    \sim e^{ \frac{\pi}{2} \xi \pm \frac{\pi}{2}\xi }.
\end{align}
The amplification is controlled by the parameter $\xi$, which depends on the inflaton velocity, $\phi' / \mathcal{H}$.
The largest effects therefore occur near the end of inflation and during reheating when the inflaton velocity is largest.
Modes with wave numbers in the band $1 \lesssim k/\mathcal{H} \lesssim 2 \xi$ are significantly enhanced by the axion.
Shortly after their production during reheating these modes can rescatter, generating the subdominant helicity as well as inflaton quanta (see \cref{fig:chaotic-phi-fluc-energy-vs-gauge}).
We use numerical simulations to capture these nonlinear dynamics in later sections.


\subsection{Gravitational waves and \texorpdfstring{$\Delta N_\mathrm{eff}$}{delta N effective}}\label{sec:gw-neff}

The tachyonic production of gauge-field modes leads to large anisotropic stresses which results in the copious production of gravitational waves~\cite{Adshead:2018doq}.
These gravitational waves form a stochastic background with frequencies near the Hubble scale during reheating, corresponding to $f \sim 10^9$ Hz today\footnote{Note that, in an abuse of notation, we use $f$ to denote both the axion decay constant as well as the frequency of the gravitational waves.} for inflation occurring near the scale of grand unification, $\Lambda \sim 10^{16} \, \mathrm{GeV}$.

Gravitational waves with wavelengths that are much shorter than the Hubble scale contribute energy density that gravitates like radiation.
During preheating, significant energy density can be deposited into subhorizon gravitational waves which, as in Ref.~\cite{Adshead:2018doq}, may be constrained by CMB and Big Bang Nucleosynthesis measurements of the total radiation density in species beyond the standard model, $\Delta N_\mathrm{eff} = N_\mathrm{eff}- 3.046$.

The net energy density in gravitational waves is given by
\begin{align}
	\Omega_{\mathrm{gw},0} h^2
	= \int \ud \ln k \, \frac{1}{\rho_0} \frac{ \ud \rho_{\mathrm{gw},0} }{\ud \ln k}.
\end{align}
By conservatively assuming that all extra radiation density present during the formation of the CMB (beyond the standard model) is composed of gravitational waves, a bound on $N_\mathrm{eff}$ directly constrains $\Omega_{\mathrm{gw},0} h^2$ via~\cite{Maggiore:1999vm}
\begin{align}
	\frac{ \Omega_{\mathrm{gw},0} h^2}{\Omega_{\gamma,0} h^2} &= \frac{7}{8} \left( \frac{4}{11} \right)^{4/3} \Delta N_\mathrm{eff},
\end{align}
where the present energy density in photons is $\Omega_{\gamma,0}h^2 = 2.47 \times 10^{-5}$.
From the Planck limit, $\vert \Delta N_\mathrm{eff} \vert \lesssim 0.33$~\cite{Ade:2015xua, Aghanim:2018eyx}, we obtain a bound $\Omega_{\mathrm{gw},0} h^2 \lesssim 1.85 \times 10^{-6}$.
Next-generation CMB experiments (e.g., CMB-S4) will probe $\Delta N_\mathrm{eff} \leq 0.03$ at $1\sigma$ and $\Delta N_\mathrm{eff} \leq 0.06$ at $2\sigma$~\cite{Abazajian:2019eic}, improving the upper limit by an order of magnitude to
\begin{align}
	\Omega_{\mathrm{gw},0} h^2 \lesssim 1.68 - 3.36 \times 10^{-7}.
\end{align}
A separate analysis~\cite{Pagano:2015hma} using Planck data provides a stronger constraint, $\Omega_{\mathrm{gw},0} h^2 < 1.2 \times 10^{-6}$ at $95\%$ confidence.
The same study projected that next-generation satellite missions COrE~\cite{Bouchet:2011ck} and Euclid~\cite{Laureijs:2011gra} will place $2\sigma$ bounds of $\Delta N_\mathrm{eff} < 0.013$ or $\Omega_{\mathrm{gw},0} h^2 < 7.6 \times 10^{-8}$.
In the results below, we explore the relationship between gravitational wave production at these levels and the preheating process itself.

The first direct detection of gravitational waves by LIGO~\cite{Abbott:2016blz} prompts consideration of direct gravitational wave detectors as alternative probes of stochastic gravitational wave backgrounds.
In particular, the presently operating LIGO~\cite{TheLIGOScientific:2014jea} and VIRGO~\cite{TheVirgo:2014hva} detectors (with KAGRA~\cite{Somiya:2011np} expected to join this network of detectors in the near future), as well as future missions like LISA~\cite{Audley:2017drz} and the Einstein Telescope~\cite{Punturo:2010zz}, could detect stochastic gravitational wave backgrounds generated in the very early universe.
In order to detect a stochastic gravitational wave background with a direct gravitational wave detector, the corresponding signal must have a sufficiently large signal-to-noise ratio (SNR), given by~\cite{Allen:1996vm, Maggiore:1999vm}
\begin{align}
    \mathrm{SNR}^2
    &= T \int_{f_{\min}}^{f_{\max}} \ud f \, \left(\frac{\Omega_\mathrm{gw}}{\Omega_s} \right)^2\\
    &\equiv T \int_{f_{\min}}^{f_{\max}} \ud f \, \left[ \frac{\Omega_\mathrm{gw}}{4 \pi^2 S_n f^3 / (3 H_0^2)} \right]^2,
\end{align}
where $\Omega_s$ is used to denote the detector sensitivity (expressed as a cosmological parameter), $S_n$ is the strain sensitivity of the instrument, $T$ denotes the observation time, and $f_{\min}$ and $f_{\max}$ are, respectively, the minimal and maximal frequencies to which the instrument is sensitive.
The fact that $\Omega_s$ is proportional to $f^3$ (times the strain) makes direct detection of high frequency gravitational waves difficult; however, if inflation occurred at a lower scale, therefore producing a lower-frequency signal, direct detection is possible~\cite{Easther:2006vd, Easther:2007vj}.
At the same time, current methods (see, for example,~\cite{Akutsu:2008qv,Goryachev:2014yra,Arvanitaki:2012cn}) for the direct detection of high-frequency gravitational waves are unfortunately not expected to be sensitive enough to place interesting bounds on stochastic backgrounds of cosmological gravitational waves.


\section{Numerical methods}\label{sec:numerical-methods}

Our numerical approach is very similar to that employed in Ref.~\cite{Adshead:2018doq}.
We numerically integrate the classical equations of motion of the inflaton, \cref{eq:phi-eom}, and the gauge fields, \cref{eqn:gauge-eom}, in an FLRW background governed by \cref{friedmann_1,friedmann_2}.
Specifically, we discretize the evolution equations onto a three-dimensional (3D), regularly-spaced grid with periodic boundary conditions and step the coupled system of equations through time using the fourth-order Runge-Kutta method.
Spacetime expansion is implemented self-consistently by computing the current energy density and pressure, \cref{rho,pressure}, averaged over the simulation volume.

Numerical treatments of gauge fields, compared to those for scalar fields, require particular care.
Specifically, numerical evolutions of gauge fields must be stable with respect to their constraints.
These constraints are Gauss's law, \cref{gauss-law}, and the prescribed gauge condition (in our case Lorenz gauge, $\partial_\mu A^\mu = 0$).

One numerical method for evolving gauge theories is lattice gauge theory, which recasts the gauge fields as a system of ``link variables'' representing the discrete connection between adjacent lattice sites.
The advantage of this method is that discrete gauge invariance is an exact symmetry of the discrete system.
The system of dynamical equations of motion that results from the direct variation of the discrete lattice action yields an evolution scheme which preserves the gauge condition and (degree of satisfaction of) Gauss's law.
However, as noted by~\cite{Cuissa:2018oiw,Moore:1996wn}, an appropriately gauge-invariant representation of the axial coupling term $\tilde{F}_{\mu\nu} F^{\mu\nu}$ renders the equations of motion implicit, requiring a computationally expensive iterative solution technique.

We take an alternative approach and evolve the Euler-Lagrange equations of the continuum theory for all of the components of the gauge field $A_\mu$ in Lorenz gauge.
That is, rather than evolve the evolution equations of a discretized theory, we numerically integrate the equations of motion of the continuum theory by discretizing the dynamical equations themselves (as often employed for scalar fields).
While gauge invariance is not exact in this case, the critical property required for robust results is \textit{stability}.
One can recast the equations of motion such that constraint violations, while dynamical, remain bounded in time~\cite{Knapp:2002fm,Baumgarte:2010:NRS:2019374}.
Doing so requires evolving additional, redundant degrees of freedom, which in our case means evolving all four components of the gauge potential,
\begin{align}
    A_\nu''
    &= \nabla^2 A_{\nu}
    + \eta_{\beta \nu} \frac{\alpha}{f} \partial_{\alpha} \phi\left(\frac{1}{2} \varepsilon^{\alpha \beta \rho \sigma} F_{\rho \sigma}\right),
\end{align}
obtained by applying the Lorenz gauge choice, $\partial_\mu A^\mu = 0$, to \cref{field-tensor-eom}.
The results of~\cite{Deskins:2013dwa,Adshead:2015pva} demonstrate the stability of the satisfaction of the gauge constraint under both the axial coupling considered here and a dilatonic coupling to the gauge-field kinetic term.
Recent work used lattice-gauge-theory--based simulations employing an iterative scheme~\cite{Cuissa:2018oiw}, the results of which reproduce those of Ref.~\cite{Adshead:2015pva} (and those presented in \cref{sec:dependence-inflationary-scale} below).
In addition, in Ref.~\cite{Adshead:2017xll} we compared simulations of preheating into three U(1) gauge fields using the approach detailed above as well as preheating into a set of SU(2) gauge fields (with the gauge self-coupling tuned so that internal interactions are negligible) using lattice gauge theory, finding near-perfect agreement between the two methods.

There are two differences in our numerical implementation relative to that in Ref.~\cite{Adshead:2018doq}.
First, we have improved our procedure for obtaining the power spectra of the gauge fields at the end of inflation.
To capture the tachyonic enhancement of (one polarization of) the gauge fields \textit{during} inflation, we numerically evolve the background spacetime, the homogeneous mode of the inflaton, and the linearized equations of motion of the gauge-field polarizations.
In contrast to Refs.~\cite{Adshead:2018doq,Adshead:2015pva}, here we more carefully account for the backreaction of the gauge fields onto the background quantities.
At the range of couplings we explore, the approximations used to derive the analytic results \cref{E2B2_analytic,EdotB_analytic} (that $\vert \xi \vert \gtrsim 4$ and is constant) are inaccurate during the final few $e$-folds of inflation.
While these approximations are valid earlier during inflation and at larger couplings (the regimes for which \cref{E2B2_analytic,EdotB_analytic} were developed~\cite{Anber:2009ua,Jimenez:2017cdr}), they overestimate the backreaction of the gauge fields onto the background at the end of inflation.
At the largest couplings we consider here (e.g., $\alpha / f \sim 15 \, \Mpl^{-1}$ for chaotic inflation), these errors artificially offset the end of inflation by $\sim 0.5$ $e$-folds.
As a result, the use of \cref{E2B2_analytic,EdotB_analytic} to compute the duration of inflation contributes errors to the relationship between present-day length scales and the number of $e$-folds before the end of inflation when those scales exited the horizon.
To mitigate these errors, we perform the integrals \cref{E2_integral,B2_integral,EdotB_integral} via quadrature of the numerically integrated gauge-field fluctuations.
See \cref{sec:initial-conditions} for full details of this procedure.

Other than improving the gauge-field power spectra, our procedure for generating initial conditions is unchanged from Ref.~\cite{Adshead:2018doq}.
We initialize the lattice simulations one to two $e$-folds (depending on the model; see \cref{tab:model-params}) before the end of inflation, where the inflaton is initialized with a mean value and velocity given by the background evolution.
We seed the fluctuations in each field and its (conformal-time) velocity as independent, Gaussian-random fields by drawing, for each Fourier mode on the lattice, an amplitude from a Rayleigh distribution and a uniform random phase.
The variance of the Rayleigh distribution for each mode is set according to the power spectrum obtained from the numerical solution of the linearized equation of motion for the field fluctuations evolved through inflation.
Finally, the chosen Lorenz gauge condition is satisfied on the initial slice by setting $A_0' = 0$ and projecting the polarization fields, $A^\pm(\mathbf{k})$, onto their vector components $A_i(\mathbf{k})$, which automatically yields $\partial_i A_i = 0$.

The second difference in our numerical implementation is that, in contrast to the pseudospectral solver used in Ref.~\cite{Adshead:2018doq}, we represent spatial derivatives with finite differencing, using fourth-order--accurate centered difference stencils.
In particular, whereas in Ref.~\cite{Adshead:2018doq} we evolved the equations of motion for the tensor metric fluctuations $h_{ij}$ in Fourier space [computing and Fourier transforming the source term \cref{gw-source} and projecting with \cref{TT-project} at each time step], we here evolve the equation of motion in position space without applying the transverse-traceless projection~\cite{GarciaBellido:2007af},
\begin{align}
    \label{uij-eom}
    u_{ij}'' + 2 \mathcal{H} u_{ij}' - \partial_k \partial_k u_{ij}
    &= \frac{2}{\Mpl^2} T_{ij}.
\end{align}
Instead, we apply the transverse-traceless projection to $u_{ij}$ to obtain
\begin{align}
    h_{i j}
    &= \left( P_{i l} P_{j m} - \frac{1}{2} P_{i j} P_{l m} \right) u_{l m},
\end{align}
only when computing the gravitational wave spectrum $\Omega_\mathrm{gw}(k)$ via \cref{GWdensitySpec}.
The advantage of this procedure is that it requires no Fast Fourier Transforms at each time step, which scale poorly to distributed-memory systems.
We have verified that the axion--gauge-field dynamics and the gravitational wave spectra $\Omega_\mathrm{gw}(k)$ reproduce the results of the pseudospectral method extremely well.

The software we developed for this purpose is \textsf{pystella},\footnote{\href{https://github.com/zachjweiner/pystella}{github.com/zachjweiner/pystella}} a Python-based, MPI-parallel and GPU-accelerated code making use of \textsf{PyOpenCL}~\cite{kloeckner_pycuda_2012} and \textsf{Loo.py}~\cite{kloeckner_loopy_2014} for the generation of OpenCL code to run on GPUs.
Because \textsf{pystella} may thus run on multiple GPUs (or any architecture with OpenCL support), we evolve lattices with $N^3 = 384^3$ points, enabling reliable simulations of larger couplings $\alpha / f$ than in Ref.~\cite{Adshead:2018doq}; we also checked a representative sample of our results against simulations with $512^3$ points to show convergence.
The Friedmann constraint, \cref{friedmann_1}, is satisfied to a precision of $\mathcal{O}(10^{-3})$ or better for all simulations we present.

The inflationary models we consider exhibit different postinflationary Hubble scales relative to the oscillation timescale about the minima of their respective potentials, $m_\phi$ (defined by $m_\phi^2 = \partial^2 V / \partial \phi^2$ evaluated at the minimum of the potential).
For this reason, we choose different comoving box lengths $L$ for different models.
In all cases, we have checked our results are insensitive to the precise choice of $L$, i.e., that the simulations have converged.
We also tune the initialization time (relative to the end of inflation) based on the model, starting sufficiently early to capture any nonlinear effects at the end of inflation (but no earlier than necessary to save computational expense).
We tabulate these choices, as well as other relevant parameters for each model, in \cref{tab:model-params}.
We use a time step $\Delta \tau = \Delta x / 10 = L/N/10$ in all cases and set the scale factor $a = 1$ at the end of inflation, when $\epsilon_H \equiv - \dot{H} / H^2 = 1$.
\begin{table*}[ht]
    \centering
    \caption{The specific parameters chosen for each inflationary model under consideration: the effective inflaton mass, the simulation box length, the simulation start time in terms of the number of $e$-folds relative to the end of inflation, the Hubble parameter at the end of inflation $H_e$, the ratio of the lattice's infrared cutoff to the comoving Hubble scale at the end of inflation, equal to $(2 \pi / L) / \mathcal{H}_e$, and the energy scale at the end of inflation.
        }\label{tab:model-params}
    \begin{tabular*}{\textwidth}[t]{l @{\extracolsep{\fill}} cccccccc}
        \toprule
        Model &             $m_\phi / \Mpl$ &           $L m_\phi$ &    $N_0$ &
        $H_e / m_\phi$ & $k_\mathrm{IR} / \mathcal{H}_e$ &  $\sqrt[4]{\rho_e} / \Mpl$ &
        $n_s$     & $r$
        \\ \midrule
        Chaotic ($n=2$) & $6.16 \times 10^{-6}$ & 15 & $-2$ & 0.51 & 0.82 & $2.3 \times 10^{-3}$ & 0.966 & 0.13
        \\
        Starobinsky ($v = 10 \Mpl/3$) & $1.06 \times 10^{-5}$ & 20 & $-2$ & 0.37 & 0.85 & $2.6 \times 10^{-3}$ & 0.969 & 0.016
        \\
        Monodromy ($\phi_c = \Mpl /10$) & $4.66 \times 10^{-5}$ & 50 & $-2$ & 0.15 & 0.84 & $3.5 \times 10^{-3}$ & 0.975 & 0.067
        \\
        Hilltop ($p = 3$, $v = M_\mathrm{pl}$) & $1.12 \times 10^{-7}$ & 20 & $-1$ & 0.11 & 2.7 & $1.5 \times 10^{-4}$ & 0.932 & $2.0 \times 10^{-8}$
        \\
        Hilltop ($p = 4$, $v = M_\mathrm{pl}$) & $8.39 \times 10^{-7}$ & 20 & $-1$ & 0.088 & 3.6 & $3.6 \times 10^{-4}$ & 0.949 & $6.4 \times 10^{-7}$
        \\
        Hilltop ($p = 4$, $v = 2 M_\mathrm{pl}$) & $1.60 \times 10^{-6}$ & 20 & $-1$ & 0.15 & 2.1 & $6.5 \times 10^{-4}$ & 0.949 & $9.8 \times 10^{-6}$
        \\
        Hilltop ($p = 4$, $v = 4 M_\mathrm{pl}$) & $3.06 \times 10^{-6}$ & 20 & $-2$ & 0.24 & 1.3 & $1.1 \times 10^{-3}$ & 0.951 & $1.4 \times 10^{-4}$
        \\
        Hilltop ($p = 4$, $v = 8 M_\mathrm{pl}$) & $5.33 \times 10^{-6}$ & 20 & $-2$ & 0.33 & 0.95 & $1.7 \times 10^{-3}$ & 0.955 & $1.7 \times 10^{-3}$
        \\
        D-brane $(p=2, v=\Mpl/2)$ & $4.90 \times 10^{-5}$ & 40 & $-1$ & 0.073 & 2.1 & $2.5 \times 10^{-3}$ & 0.975 & $2.2 \times 10^{-3}$
        \\
        Natural ($v = \sqrt{8 \pi} \Mpl$) & $4.85 \times 10^{-6}$ & 15 & $-2$ & 0.50 & 0.84 & $2.1 \times 10^{-3}$ & 0.952 & 0.033
        \\
        \bottomrule
    \end{tabular*}
\end{table*}


\section{Analytic estimates}\label{sec:analytic-estimates}

To establish expectations for the effects of the inflationary model on preheating and gravitational wave production, we begin with some analytical estimates.
While the linear theory reviewed in \cref{App:eomsandlinalaysis} is not valid during preheating, it can be used to gain intuition for the scaling of the backreaction on the expansion rate and on the motion of the inflaton during the initial phase of preheating which is typically the most violent.
We additionally use a ``rule of thumb'' for stochastic gravitational wave production from cosmological processes~\cite{Giblin:2014gra} to make contact between the characteristics of the inflationary potential and the efficiency of gravitational wave production from preheating.


\subsection{Efficiency of preheating}\label{sec:analytic-estimates-efficiency}

From the Friedmann equation, \cref{friedmann_1}, by making use of the approximation for the energy density in the gauge fields, \cref{EdotB_analytic}, we can derive the approximation
\begin{align}
    \label{eq:approx_hubble}
    \begin{split}
        \frac{H^2}{\Mpl^2 }
        &= \frac{1}{\left( 3 - \epsilon_{\phi} \right)} \frac{V}{\Mpl^4 } \left[ 1 + \frac{V}{\Mpl^4} \frac{ 1.4 \times 10^{-4} \, e^{2 \pi \xi} }{\xi^3 \left( 3 - \epsilon_{\phi} \right)^2 } \right] \\
        &\hphantom{={}}
        + \mathcal{O}\left(\frac{V}{\Mpl^4}\right)^3,
    \end{split}
\end{align}
where we introduce the slow-roll parameters
\begin{align}\label{eqn:epsSR}
    \epsilon_\phi \equiv \frac{\dot{\phi}^2}{2 \Mpl^2 H^2} , \quad \epsilon_{H} \equiv -\frac{\dot{H}}{H^2}.
\end{align}
In standard single-field slow-roll inflation, the Einstein equations imply that $\epsilon_{\phi} = \epsilon_{H}$.
However, in the regime of strong gauge-field backreaction, this relation does not hold.
\Cref{eq:approx_hubble} quantifies the contribution of the gauge fields to the Hubble rate relative to the contribution from the inflaton.
As such, it provides a rough estimate of the value of $\xi$ (and therefore $\phi'/\mathcal{H}$, once $\alpha/f$ is fixed) required in order for the gauge-field contribution to the energy density to be comparable with the inflaton's.

Beyond modifying the value of the Hubble parameter as in \cref{eq:approx_hubble}, the gauge fields induce a new friction term in the equation of motion of the inflaton.
Substituting the spatially averaged interaction, \cref{EdotB_analytic}, into the equation of motion for the inflaton, \cref{eq:phi-eom}, while making use of the lowest order approximation of \cref{eq:approx_hubble} and assuming $\phi''/\mathcal{H}^2 \ll \Mpl$, we can derive the usual slow-roll condition, now including the effect of the gauge-field backreaction,
\begin{align}
    \label{friction_comparison}
    \begin{split}
    \Mpl^2 \frac{V_{, \phi}}{V}
    &\simeq -\frac{\dot{\phi}}{H}\left[ 1 + \frac{V}{\Mpl^4} \frac{ 1.4 \times 10^{-4} \, e^{2 \pi \xi} }{\xi^3 \left( 3 - \epsilon_{\phi} \right)^2 } \right] \\
    &\hphantom{={}}
    + 2.4 \times 10^{-4} \frac{\alpha}{f} \frac{V}{\Mpl^2} \frac{ e^{2 \pi \xi} }{\xi^4 \left( 3 - \epsilon_\phi \right)^2}.
    \end{split}
\end{align}
Note that the gauge-field backreaction changes the relationship between the slope of the potential relative to its magnitude and so also the slow-roll parameter in \cref{eqn:epsSR}.
In this scenario, gauge-field--induced friction provides an additional mechanism to enforce the flatness of the potential required for slow-roll inflation~\cite{Anber:2009ua,Barnaby:2011qe,Domcke:2016bkh}.

There are two effects at play in \cref{friction_comparison}.
The first is that the backreaction changes the Hubble rate, increasing Hubble friction (on the inflaton), while the second is the backreaction of the $\mathbf{E}\cdot\mathbf{B}$ term.
The impact of these two effects can be understood by comparing \cref{friction_comparison} with \cref{eq:approx_hubble}.
As the same ($V$-dependent) term is compared with unity (for the backreaction on $H$) and with $\sqrt{2 \epsilon_{\phi}}$ (for the backreaction on the equation of motion of the inflaton) respectively, the gauge-field--induced modification of the Hubble parameter is higher order (in $V / \Mpl^4$) than the new friction term in the equation of motion of the inflaton.
In addition, while both the gradient of the potential and the (lowest order expression of the) Hubble friction term in \cref{friction_comparison} do not depend explicitly on the scale of $V$, the gauge-field friction depends linearly on the inflationary energy scale.
As a consequence, low-scale models are expected to require a larger value of $\xi$ in order for the gauge fields to become important.

Finally, while $V$ is nearly constant (which is the case during inflation), the gauge-field--induced backreaction grows exponentially with $\xi$.
However, if $V$ is not constant (as is the case during preheating where it decreases as fast as, or faster than, $e^{-2\pi \xi}$) then the gauge field induced backreaction may either never become relevant or be shut off.
Such a possibility is realized in potentials without minima, such as a $\tanh(\phi/M)$ shape.
We leave investigations of this class of potentials to future work.

To set expectations for the range of couplings $\alpha / f$ relevant for preheating in a given model, we make use of \cref{eq:approx_hubble} (to leading order in slow roll) to obtain scaling relations between $\alpha / f$ and the Hubble scale at the onset of preheating (and so the parameters of the model).
We quantify efficiency in terms of the fraction of the Universe's energy residing in the gauge fields,
\begin{align}
    \frac{\rho_\mathrm{gauge}}{\rho}
    &\approx
    \frac{1.4 \times 10^{-4}}{3} \frac{(H / \Mpl)^2}{\xi^3} e^{2 \pi \xi}.
\end{align}
In the regime of this approximation's validity, we may ask how much one must tune $\xi$ (or $\alpha / f$) to compensate for a reduction in $H / \Mpl$ in order to keep $\rho_\mathrm{gauge} / \rho$ fixed.
Considering two sets of parameters $(H_1, \xi_1)$ and $(H_2, \xi_2)$, this amounts to solving the nonlinear equation
\begin{align}\label{fgauge_scaling}
    \frac{\rho_{\mathrm{gauge}, 1}}{\rho_{\mathrm{gauge}, 2}}
    = \left( \frac{H_1}{H_2} \right)^2 \left( \frac{\xi_2}{\xi_1} \right)^3 e^{2 \pi (\xi_1 - \xi_2)} = 1
\end{align}
for $\xi_2$ in terms of $H_1 / H_2$ and $\xi_1$.
When $\xi_1$ and $\xi_2$ are both small (and so $\xi_1 - \xi_2$ is as well), we solve
\begin{align}
    \left( \frac{H_1}{H_2} \right)^2 \left( \frac{\xi_2}{\xi_1} \right)^3
    \approx 1,
\end{align}
yielding $\xi_2 = \xi_1 (H_2 / H_1)^{2/3}$.
Alternatively, if $\xi_1$ and $\xi_2$ are both large,
\begin{align}\label{large-xi-scaling}
    \left( \frac{H_1}{H_2} \right)^2 e^{2 \pi (\xi_1 - \xi_2)}
    \approx 1,
\end{align}
in which case $\xi_2 = \xi_1 + \ln (H_1 / H_2) / \pi$.
From this we read that the coupling must increase by an additive amount proportional to the logarithm of the ratio of Hubble scales.
In our analysis below we demonstrate that this scaling relation is sufficiently accurate for our estimates.

References~\cite{Adshead:2015pva,Adshead:2016iae,Adshead:2018doq} demonstrate that in chaotic inflation models with $m_\phi \approx 10^{-6}$, the threshold value of $\alpha / f$ for which preheating is complete (i.e., $\max \rho_\mathrm{gauge} / \rho \sim 80\%$) is $9 \, \Mpl^{-1}$.
With this as a baseline, we apply \cref{fgauge_scaling} to estimate how this threshold scales as we change the energy scale of inflation (and, later, the inflaton model itself).
Consider the simplest case of a quadratic potential, an approximation useful for all of our models during the preheating phase,
\begin{align}
    V(\phi) = \frac{1}{2} m_\phi^2 \phi^2.
\end{align}
In this case, the Hubble parameter scales as $H \sim m_\phi$, since at the end of inflation
\begin{align}
    H^2 = \frac{\rho}{3 \Mpl^2}
    \approx \frac{V(\phi)}{3 \Mpl^2}
    \sim m_\phi^2.
\end{align}
In terms of the coupling, \cref{large-xi-scaling} tells us that, for some other mass $m_{\phi, 2}$, we require
\begin{align}\label{scaling-in-terms-of-af}
    \frac{\alpha_2}{f} = \frac{\alpha_1}{f} + \frac{1}{\pi \abs{\partial \phi / \partial N}} \ln \left( \frac{m_{\phi, 1}}{m_{\phi, 2}} \right)
\end{align}
for comparable preheating efficiency.
Since we are interested in the efficiency of preheating (rather than the inflationary production of gauge bosons), in this expression we evaluate $\partial \phi / \partial N$ at the end of inflation, $N = 0$.
At the end of chaotic inflation, $\partial \phi / \partial N \approx 1.4 \, \Mpl$, in which case \cref{scaling-in-terms-of-af} reduces to
\begin{align}\label{approximate-scaling-chaotic}
    \frac{\alpha_2}{f} = \frac{\alpha_1}{f} + \frac{1.1}{\Mpl} \log_{10} \left( \frac{m_{\phi, 1}}{m_{\phi, 2}} \right).
\end{align}
Thus, for comparable preheating efficiency we expect to need to increase the axion-gauge coupling $\alpha / f$ by $\sim \, \Mpl^{-1}$ for each order of magnitude we reduce $m_\phi$.


\subsection{Gravitational wave production and the ``rule of thumb''}\label{sec:rule-of-thumb}

While regimes of highly efficient preheating exist for all models (as we show in \cref{sec:dependence-potential-shape}), the structure of resonance varies from model to model.
To understand the relationship between the scales at which gravitational waves are produced and their resulting amplitude, we use the ``rule of thumb'' developed by the authors of Ref.~\cite{Giblin:2014gra} to estimate the stochastic gravitational wave production from cosmological processes.
By approximating the source as a Gaussian of width $\sigma$ peaked at wave number $k_\ast$, Ref.~\cite{Giblin:2014gra} estimates the peak amplitude as
\begin{align}\label{thumb}
    \Omega_{\mathrm{gw}, 0}
    \approx 2.3 \times 10^{-4} \alpha^2 \beta w^2 \frac{k_\ast}{\sigma} \left(\frac{H_\ast}{k_\ast}\right)^2,
\end{align}
where $\alpha$ is the fraction of the energy in the gravitational wave source relative to the Universe's total energy density at that time, $\beta$ encodes the anisotropy of the source, and $w$ is the equation of state of the Universe at that time.

Observe that $\Omega_{\mathrm{gw}, 0} h^2$ decreases quadratically with the ratio of the peak wave number $k_\ast$ to the Hubble parameter at that time, $H_\ast$.
The oscillation frequency of the inflaton background (i.e., its effective mass) sets the scales of efficient resonance, while $\xi \sim \alpha / f$ controls how far inside the horizon tachyonic resonance occurs.
The former depends on the shape of the potential, while the latter can depend on the inflationary scale via the arguments of \cref{sec:analytic-estimates-efficiency}: lower inflationary scales require larger couplings for complete preheating.
To quantify the effect of the potential shape, note that the effective inflaton mass, defined by $m_\phi^2 \equiv \partial^2 V / \partial \phi^2$ evaluated at the minimum of the potential, sets the scales of interest during preheating.
As such, $k_\ast / H_\ast \sim m_\phi / H_\ast \sim \Mpl \sqrt{V'' / V}$---i.e., the shape of the potential (about its minimum) determines the scales at which preheating occurs, which in turn affects the size of the gravitational wave signal according to \cref{thumb}.


\section{Numerical results}


\subsection{Dependence on inflationary scale}\label{sec:dependence-inflationary-scale}

In this section, we present simulations detailing the effect of the energy scale of inflation on the efficiency of preheating and the subsequent generation of gravitational waves.
For this analysis we fix a chaotic inflationary potential while tuning $m_\phi$ to study different energy scales of inflation (i.e., for the time being, we ignore the fact that $m_\phi$ should be chosen to fit the normalization of the scalar power spectrum during inflation).
This simplification allows us to separate the effect of the scale of the potential from that of its shape, which we consider in \cref{sec:dependence-potential-shape}.
Further, preheating studies are often restricted to chaotic inflationary models, a choice justified because preheating probes the inflaton's oscillation about the minimum of its potential, which is quadratic to leading order.

\begin{figure*}[t]
    \centering
    \includegraphics[width=\textwidth]{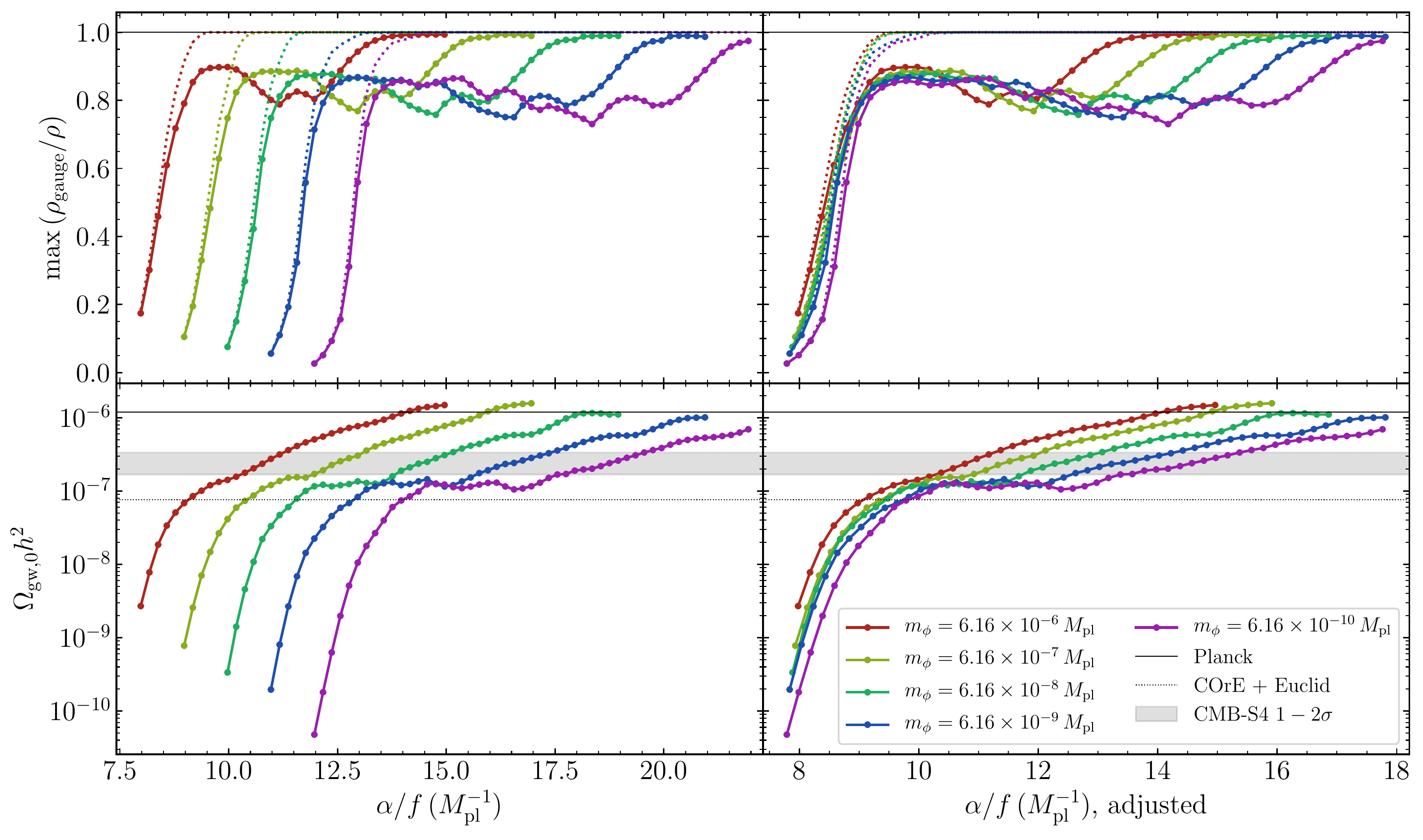}
    \caption{
        Left panels: Preheating efficiency, quantified by the maximum $\rho_\mathrm{gauge} / \rho$ over the simulation (top panels), and the total fractional energy in gravitational waves today, $\Omega_{\mathrm{gw}, 0} h^2$ (bottom panel), as functions of axion-gauge coupling $\alpha / f$.
        Each curve fixes a quadratic potential and corresponds to various values of $m_\phi$, decreasing by factors of 10 from $m_\phi = 6.16 \times 10^{-6} \, \Mpl$ (red) to $m_\phi = 6.16 \times 10^{-10} \, \Mpl$ (purple).
        The dashed lines (which follow the same color scheme) correspond to the maximum total energy in fluctuations, i.e., the maximum amount of energy in the simulation outside of the inflaton condensate.
        Lines indicating the current $\Delta N_\mathrm{eff}$ bounds on $\Omega_{\mathrm{gw}, 0} h^2$ from Planck and CMB-S4 from Ref.~\cite{Pagano:2015hma} are plotted in solid and dashed black, respectively, while the region between CMB-S4's $1 \sigma$ and $2 \sigma$ projections~\cite{Abazajian:2019eic} is shaded grey.
        Right panels: Same, but with the horizontal axis adjusted according to \cref{scaling-in-terms-of-af}.
        That is, the horizontal axis corresponds to the actual coupling $\alpha / f$ shifted to the coupling that would be required for the same preheating efficiency for $m_\phi = 6.16 \times 10^{-6} \, \Mpl$ (the value which fits the amplitude of the scalar power spectrum), as predicted by the analytic estimate \cref{scaling-in-terms-of-af}.
    }\label{fig:gw-money-vs-mphi}
\end{figure*}

We first seek to determine how the axion-gauge coupling $\alpha / f$ must be tuned in order to achieve complete preheating as we lower the inflationary scale and to evaluate the accuracy of the analytic estimates made in \cref{sec:analytic-estimates-efficiency}.
In \cref{fig:gw-money-vs-mphi} we depict the relationship between the efficiency of preheating and the net gravitational wave production across five decades of $m_\phi$.
Because larger couplings result in power transfer to modes with larger momentum, for values of $m_\phi < 10^{-6} \Mpl$ we use a box length $L = 7.5 \, m_\phi^{-1}$ in order to ensure sufficient short-wavelength resolution in all cases.
We immediately observe that \cref{approximate-scaling-chaotic} does indeed accurately predict the range of couplings for which preheating becomes efficient and even complete ($\rho_\mathrm{gauge} / \rho \gtrsim 80\%$).
It is reassuring that this scaling argument---derived from linearized approximations---is applicable and that it breaks down in the large-coupling limit when the approximations are least valid.
To visualize this scaling, in the right panel of \cref{fig:gw-money-vs-mphi} we plot the preheating efficiency and net gravitational wave production as functions of an ``adjusted'' coupling, i.e., the value of $\alpha / f$ that would be required to obtain the same efficiency were $m_\phi = 6.16 \times 10^{-6} \, \Mpl$, as predicted by the analytic estimate \cref{scaling-in-terms-of-af}.

\begin{figure*}[t]
    \centering
    \includegraphics[width=\textwidth]{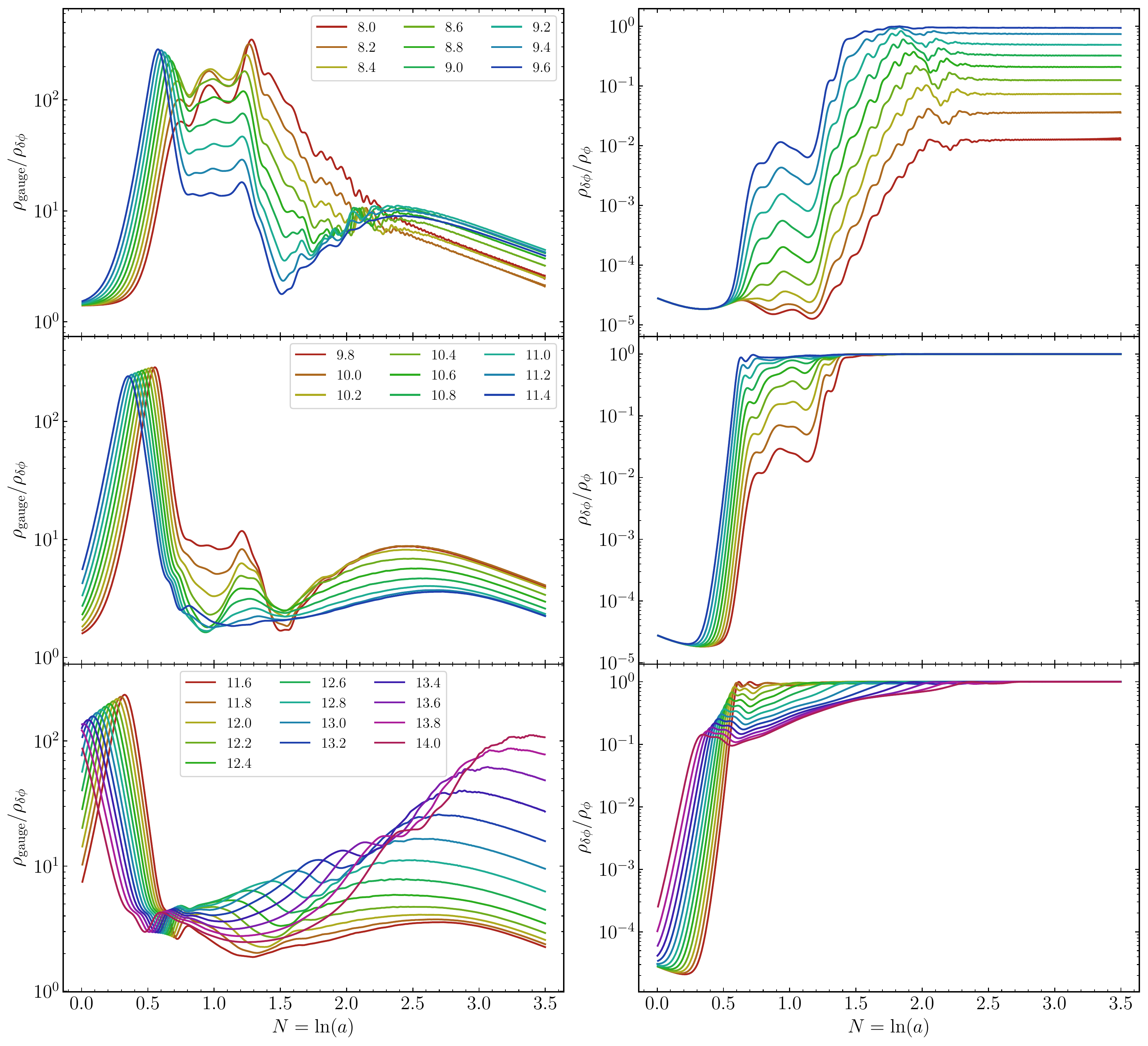}
    \caption{
        Left: Ratio of the energy in gauge fields, $\rho_\mathrm{gauge}$, to the energy in inflaton fluctuations, $\rho_{\delta \phi}$ (defined in \cref{def-rho-delta-phi}), as a function of $e$-folds after inflation, $N = \ln a$.
        Right: Fragmentation of the inflaton, measured by $\rho_{\delta \phi} / \rho_\phi$.
        Each row depicts a different coupling regime, with couplings $\alpha / f$ labeled in the legend of the left column (in units of $\Mpl^{-1}$).
        The displayed simulations fix a chaotic inflationary potential with $m_\phi = 6.16 \times 10^{-6} \, \Mpl$ to fit the amplitude of the scalar power spectrum.
    }\label{fig:chaotic-phi-fluc-energy-vs-gauge}
\end{figure*}

The efficiency of gauge field production becomes more complicated at larger couplings where nonlinear processes become important.
After a regime of $\alpha / f$ in which preheating remains comparably efficient (which, for $m_\phi = 6.16 \times 10^{-6} \, \Mpl$, corresponds to $9 \, \Mpl^{-1} \lesssim \alpha / f \lesssim 10.4 \, \Mpl^{-1}$), $\rho_\mathrm{gauge} / \rho$ begins to decrease.
At even stronger coupling, energy transfer to the gauge fields continues, gradually increasing $\rho_\mathrm{gauge} / \rho$ to $100\%$.
Regardless, the inflaton condensate is totally depleted, as shown by the dotted lines in \cref{fig:gw-money-vs-mphi} which indicate the (maximum) fraction of energy in either gauge fields or fluctuations of the axion.
To examine the physics that realizes this trend, consider comparing the energy in inflaton fluctuations,
\begin{align}\label{def-rho-delta-phi}
    \rho_{\delta \phi}
    &\equiv \rho_\phi - \rho_{\left\langle \phi \right\rangle} \\
    \begin{split}
        &= \left\langle \frac{1}{2 a^2} {\phi'}^2 + \frac{1}{2 a^2} \partial_i \phi \partial_i \phi + V(\phi) \right\rangle \\
        &\hphantom{={}}
        - \left( \frac{1}{2 a^2} {\left\langle \phi \right\rangle'}^2 + V(\left\langle \phi \right\rangle) \right),
    \end{split}
\end{align}
to the energy in the gauge fields, $\rho_\mathrm{gauge}$.
In \cref{fig:chaotic-phi-fluc-energy-vs-gauge} we plot the ratio of these two quantities, $\rho_\mathrm{gauge} / \rho_{\delta \phi}$, as well as $\rho_{\delta \phi} / \rho_\phi$ which measures the degree to which the axion has fragmented.
In the top row, we observe that as preheating approaches $\sim 80\%$ efficiency (at $\alpha / f \sim 9 \, \Mpl^{-1}$), the inflaton also becomes more fragmented.
Since the axion's equation of motion is linear for a quadratic potential, any inflaton particle production must occur through backscattering from produced gauge bosons, an inherently nonlinear process.
As the coupling increases to $9 \, \Mpl^{-1} \lesssim \alpha / f \lesssim 10.2 \, \Mpl^{-1}$, the inflaton becomes totally fragmented and the energy in gauge fields relative to inflaton fluctuations does not increase.
Preheating thus remains comparably efficient in this coupling regime.
Evident in the middle row is that, as $\alpha / f$ is increased past $10.2 \, \Mpl^{-1}$ and inflaton fragmentation occurs more rapidly, the energy in fluctuations of the axion approach roughly half that of the gauge fields.
The exponential amplification of the gauge fields likewise ends earlier and earlier, as the tachyonic resonance is driven by the motion of the (now-depleted) inflaton condensate.
Thus, preheating becomes slightly less efficient, as apparent in \cref{fig:gw-money-vs-mphi}.

Increasing $\alpha / f$ past $11.6 \, \Mpl^{-1}$, we see that while the initial phase of resonance continues to end earlier and earlier, the inflaton also becomes less fragmented during this phase.
The backreaction of the gauge fields onto the axion's background dynamics now exerts a dramatic amount of friction before the axion even first crosses through the minimum of its potential.
Tachyonic amplification of the gauge fields recommences and proceeds at a slower and slower rate as the coupling increases (as this increases the gauge-field friction).
This process terminates later and later at values of $\rho_\mathrm{gauge} / \rho$ increasingly close to $100\%$, as shown in \cref{fig:gw-money-vs-mphi}, at which point the inflaton condensate is totally depleted.
Indeed, preheating is $100\%$ efficient in these cases because the gauge-field friction is strong enough to ensure that tachyonic amplification continues until preheating is complete (in contrast to the axion's oscillations allowing for backscattering effects to become important).
Note that these results (which reproduce those originally obtained in Ref.~\cite{Adshead:2015pva}) are very similar to those presented more recently in Ref.~\cite{Cuissa:2018oiw} for the case of chaotic inflation, and our analyses of these regimes are likewise similar.

\begin{figure}[t]
    \centering
    \includegraphics[width=\columnwidth]{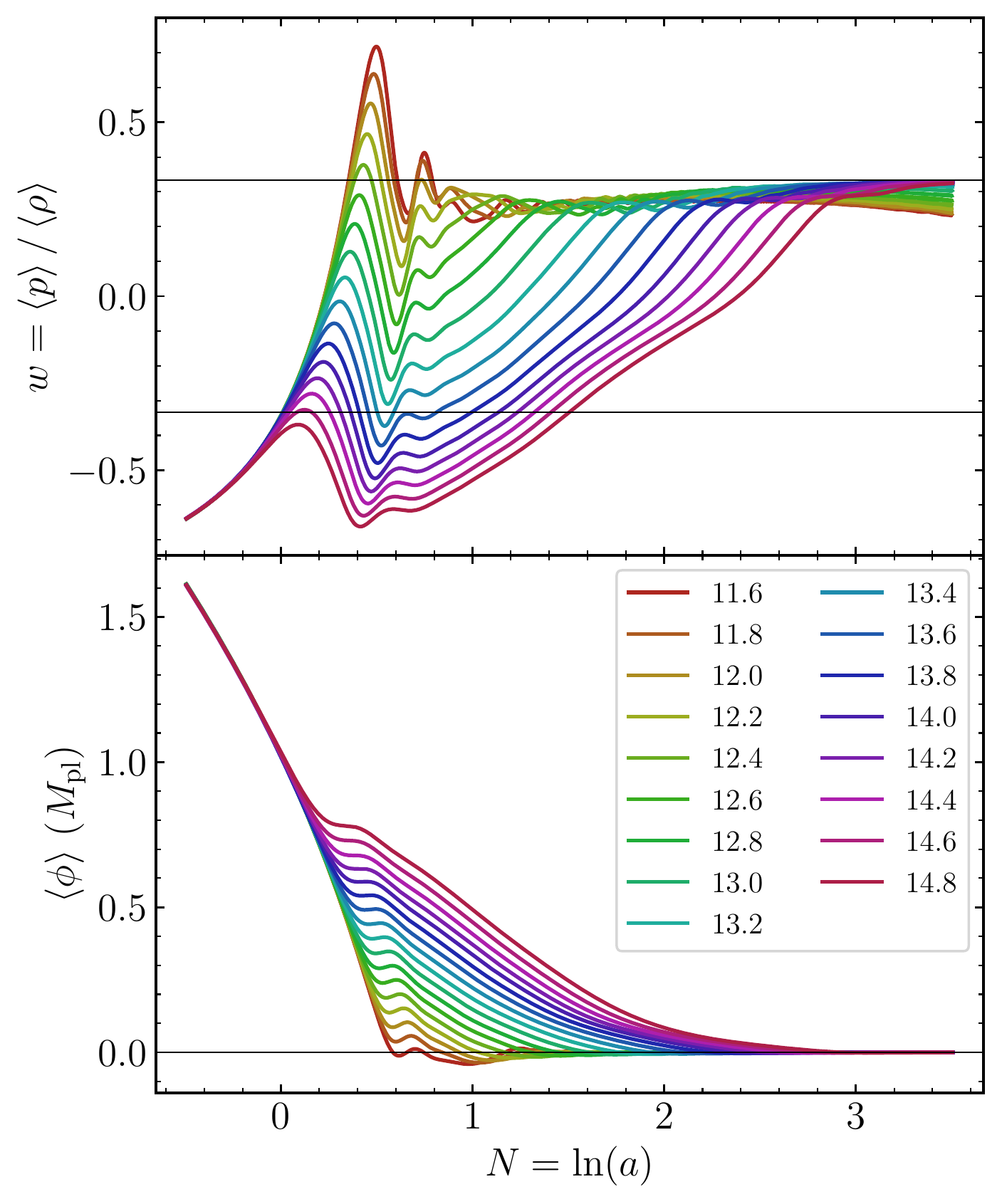}
    \caption{
        The equation of state, $w = \left\langle P \right\rangle / \left\langle \rho \right\rangle$ and the mean value of the axion field $\left\langle \phi \right\rangle$.
        Couplings range from $\alpha / f = 11.6 \, \Mpl^{-1}$ to $14.8 \, \Mpl^{-1}$ as labeled in the legend.
    }\label{fig:chaotic-phi-eos-and-phi-mean}
\end{figure}

The inflaton halts higher and higher up its potential as the gauge-field friction becomes more important, and eventually accelerated expansion ($w < - 1/3$) recommences (for $\alpha / f \gtrsim 13.2 \, \Mpl^{-1}$).
As observed in Ref.~\cite{Adshead:2016iae}, the inflaton is momentarily ``trapped,'' as depicted in \cref{fig:chaotic-phi-eos-and-phi-mean}.
Its vanishing velocity momentarily shuts off the tachyonic instability, so that the gauge fields redshift and the inflaton starts to roll again (restarting the resonant enhancement of the gauge fields).
At the highest couplings we simulate here ($\alpha / f \sim 14.8 \, \Mpl^{-1}$), the inflaton is trapped before the background stops accelerating (i.e., before $w \geq -1/3$), resembling models of inflation which achieve slow roll via the same axion--gauge-field coupling~\cite{Anber:2009ua}.

Regardless of the value of $m_\phi$, each set of simulations exhibits the same qualitative features as the coupling increases.
However, as $m_\phi$ is decreased, the range of couplings $\alpha / f$ spanned by these features broadens in a manner that cannot be described by the analytic estimates above.
The backscattering effects we posit as responsible for the dip in efficiency depend on the amplitude of axion fluctuations, which scales linearly with $m_\phi$.
Thus, in order for backscattering to be important to a comparable degree, the coupling has to be increased by a greater relative amount to compensate for the smaller amplitude of vacuum fluctuations at lower values of $m_\phi$.
From the analytical estimates presented at the beginning of this section, we expect the backreaction of the gauge fields onto the axion condensate to scale with the energy scale of the inflationary potential.
Thus, models with lower $m_\phi$ require a larger relative increase in $\alpha / f$ to enter the regime of slow tachyonic resonance due to the gauge fields' friction on the axion.

\begin{figure}[t]
    \centering
    \includegraphics[width=\columnwidth]{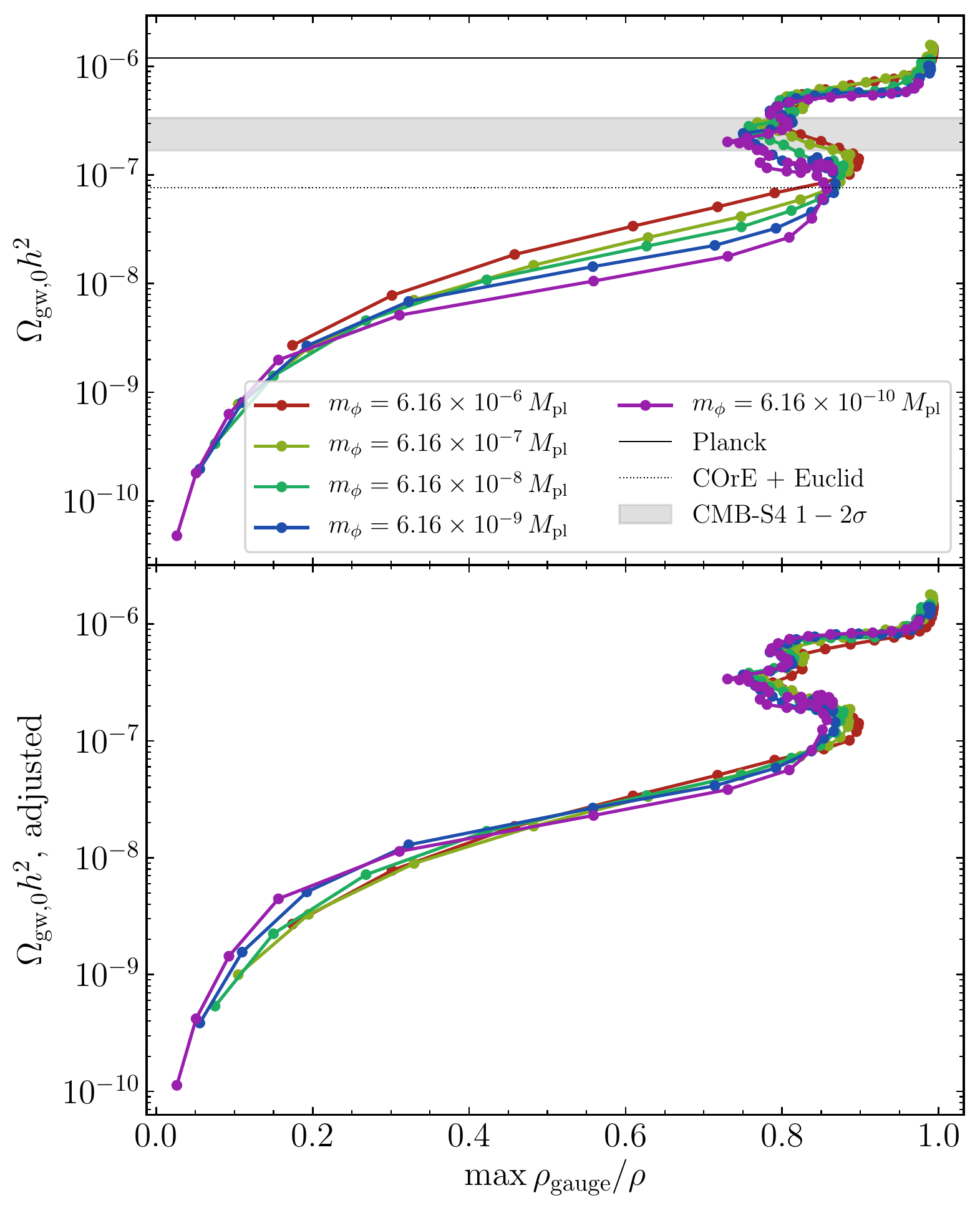}
    \caption{
        The total gravitational wave energy today versus the final preheating efficiency $\rho_\mathrm{gauge} / \rho$, plotted for a chaotic potential with varying $m_\phi$.
        The bottom panel depicts the same data, but scaled by the square of the ratio of $\alpha / f$ (for a given simulation) to the coupling required for equal preheating efficiency for $m_\phi = 6.16 \times 10^{-6} \, \Mpl$, as estimated by \cref{scaling-in-terms-of-af}.
    }\label{fig:chaotic-mphi-gw-total-vs-max-frac}
\end{figure}

The approximate universality we observe also extends to the gravitational waves produced by the dynamics of preheating.
Of note in \cref{fig:gw-money-vs-mphi} is that any scenario which exhibits (near-) complete preheating (namely, $\rho_\mathrm{gauge} / \rho \sim 80\%$) results in a net gravitational wave production that could be detected (or ruled out) by CMB-S4, as discussed in \cref{sec:gw-neff}.
For the strongest couplings we simulate, the total integrated $\Omega_{\mathrm{gw}, 0} h^2$ exceeds $10^{-6}$, which is already ruled out by Planck data~\cite{Pagano:2015hma}.
To quantify this claim, in \cref{fig:chaotic-mphi-gw-total-vs-max-frac} we scatterplot the fractional energy density in gravitational waves today versus the efficiency of preheating.
The cluster of data points around and above $\max \rho_\mathrm{gauge} / \rho \sim 0.8$ mostly lies above the CMB-S4 bound, indicating that next-generation experiments will place bounds on the axion--gauge-field coupling $\alpha / f$, regardless of $m_\phi$.

In the upper panel of \cref{fig:chaotic-mphi-gw-total-vs-max-frac} we also observe that gravitational wave production is moderately less effective at lower inflationary scales.
Returning to the arguments of \cref{sec:rule-of-thumb}, the larger couplings required for comparable preheating at lower $m_\phi$ push the resonance further inside the horizon.
As the momenta undergoing the tachyonic instability are $k / a H < \xi \sim \alpha / f$, from the rule of thumb, \cref{thumb}, we expect $\Omega_\mathrm{gw}$ to decrease with $(k_\ast / H)^2 \sim (\alpha / f)^2$.
To verify this scaling, in the lower panel of \cref{fig:chaotic-mphi-gw-total-vs-max-frac} we scale $\Omega_{\mathrm{gw}, 0} h^2$ by the squared ratio of each simulation's coupling $\alpha / f$ to its ``adjusted'' value from \cref{scaling-in-terms-of-af}.
With this multiplicative factor, the trends of $\Omega_{\mathrm{gw}, 0} h^2$ versus $\max \rho_\mathrm{gauge} / \rho$ line up closely, confirming our hypothesis.

\begin{figure*}[t]
    \centering
    \includegraphics[width=\textwidth]{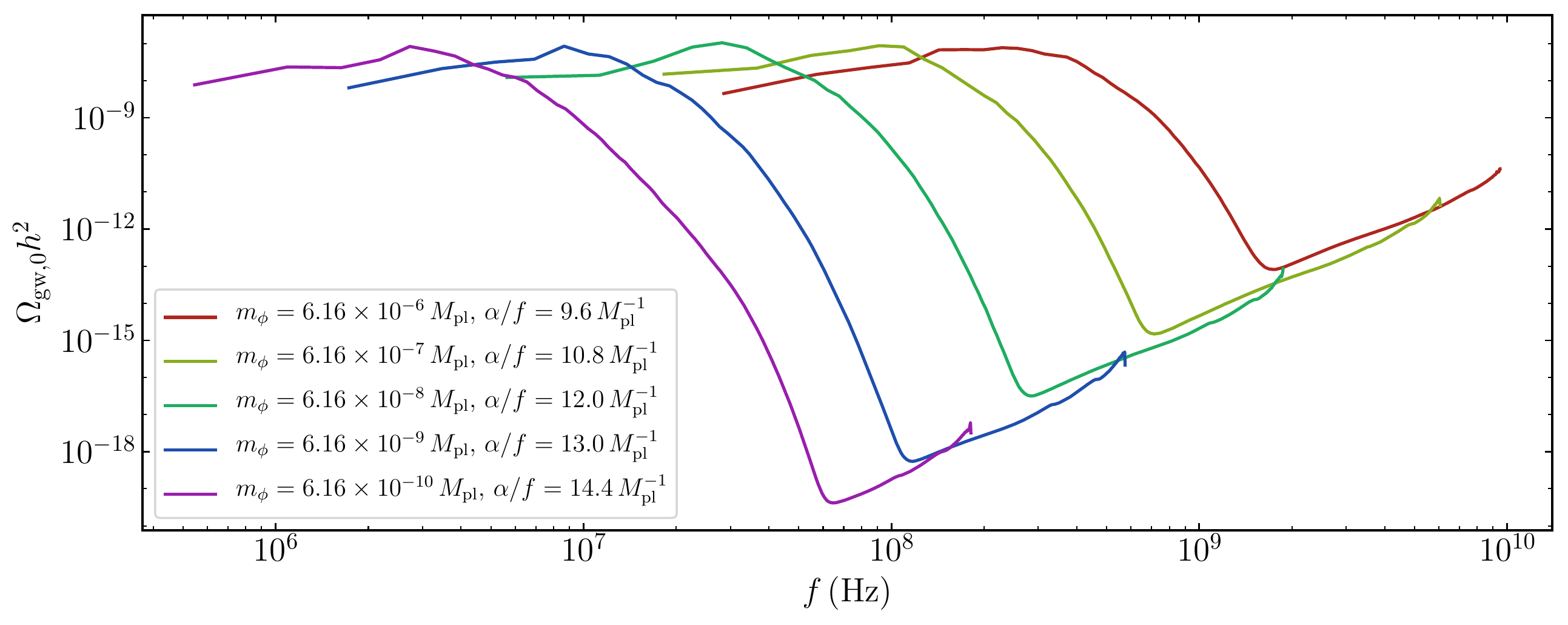}
    \caption{
        Gravitational wave spectrum observed today for various inflationary energy scales.
        Each curve corresponds to a different simulation of preheating in chaotic inflation, with coupling $\alpha / f$ and inflaton mass $m_\phi$ labeled in the legend.
        In each case, the total amount of gravitational wave production $\Omega_{\mathrm{gw},0} \sim 10^{-7}$.
    }\label{fig:gw-compare-mphi-vs-f}
\end{figure*}

The shape of the gravitational wave spectra (which is fairly generic and qualitatively comparable to that produced by tachyonic resonances) is similar for all values of $m_\phi$ (when comparing couplings which yield comparable preheating efficiency).
Plotting the signals that would be observed today in \cref{fig:gw-compare-mphi-vs-f} demonstrates this observation, and also depicts the $\sqrt{m_\phi}$ scaling of the characteristic present-day frequencies at which these signals would be observed.\footnote{Note that the ultraviolet parts of the spectra in \cref{fig:gw-compare-mphi-vs-f}, which grow as $k^4$, are not physical signals but the product of vacuum modes in the simulation.}

In general, $\Omega_{\mathrm{gw}, 0} h^2$ increases exponentially with $\alpha / f$, which is to be expected as the tachyonic resonance is exponential in $\xi \propto \alpha / f$.
The actual rate at which $\Omega_{\mathrm{gw}, 0} h^2$ increases as a function of $\exp(\alpha / f)$ falls into two regimes: the range of couplings where the actual efficiency of preheating increases exponentially with $\alpha / f$, and those for which preheating is always complete (the latter of which is slower than the former).
In the first regime the gravitational wave source (parametrized by $\alpha$ in \cref{thumb}) is growing exponentially.
In the second, while the simulations all completely transition to radiation domination, gravitational waves are continually sourced by the second phase of slow tachyonic resonance at an efficiency which still increases with the coupling strength.

Last, we note that, despite the complicated trend of $\rho_\mathrm{gauge} / \rho$ as $\alpha / f$ increases, the inflaton condensate always ends up depleted as the coupling increases past the critical value where efficient preheating is first achieved (e.g., $\alpha / f \sim 9.6 \, \Mpl^{-1}$ for $m_\phi = 6.16 \times 10^{-6} \, \Mpl$).
Thus, in these cases the end state of the simulations is always radiation domination.
However, the proportion of that radiation composed of axion fluctuations varies with coupling, as depicted in \cref{fig:chaotic-phi-fluc-energy-vs-gauge}.
Ignoring any decay channels, the axion's fluctuations redshift after preheating until their physical momenta drop below their mass, at which point they become nonrelativistic.
From this point on the axion energy density redshifts as matter.
If the axion's lifetime is sufficiently long, its energy eventually dominates over the gauge fields (which, being radiationlike, decay faster than matter).
Any deviation from an equation of state of radiation, $w \equiv p / \rho = 1/3$, suppresses the gravitational wave density observed today, $\Omega_\mathrm{gw, 0} h^2$, relative to what the transfer function \cref{gw-transfer-function} accounts for (which assumes the Universe was radiation dominated from the time of emission until matter-radiation equality).
In \cref{sec:post-preheating-dynamics} we demonstrate that Bose enhancement resulting from the larger occupation numbers from preheating ensures that perturbative decays happen quickly enough that the Universe remains radiation dominated.
As such, we expect little to no suppression of $\Omega_\mathrm{gw, 0} h^2$ relative to the values we report.


\subsection{Dependence on the shape of the potential}\label{sec:dependence-potential-shape}

We now explore the dependence of our results---in particular the amplitude of the resulting gravitational wave spectrum---on the shape of the potential during the reheating phase.
We simulate preheating in the inflationary models detailed in \cref{sec:modelsub} and discuss the extent to which the results of \cref{sec:dependence-inflationary-scale} are modified.

\begin{figure}[t]
    \centering
    \includegraphics[width=\columnwidth]{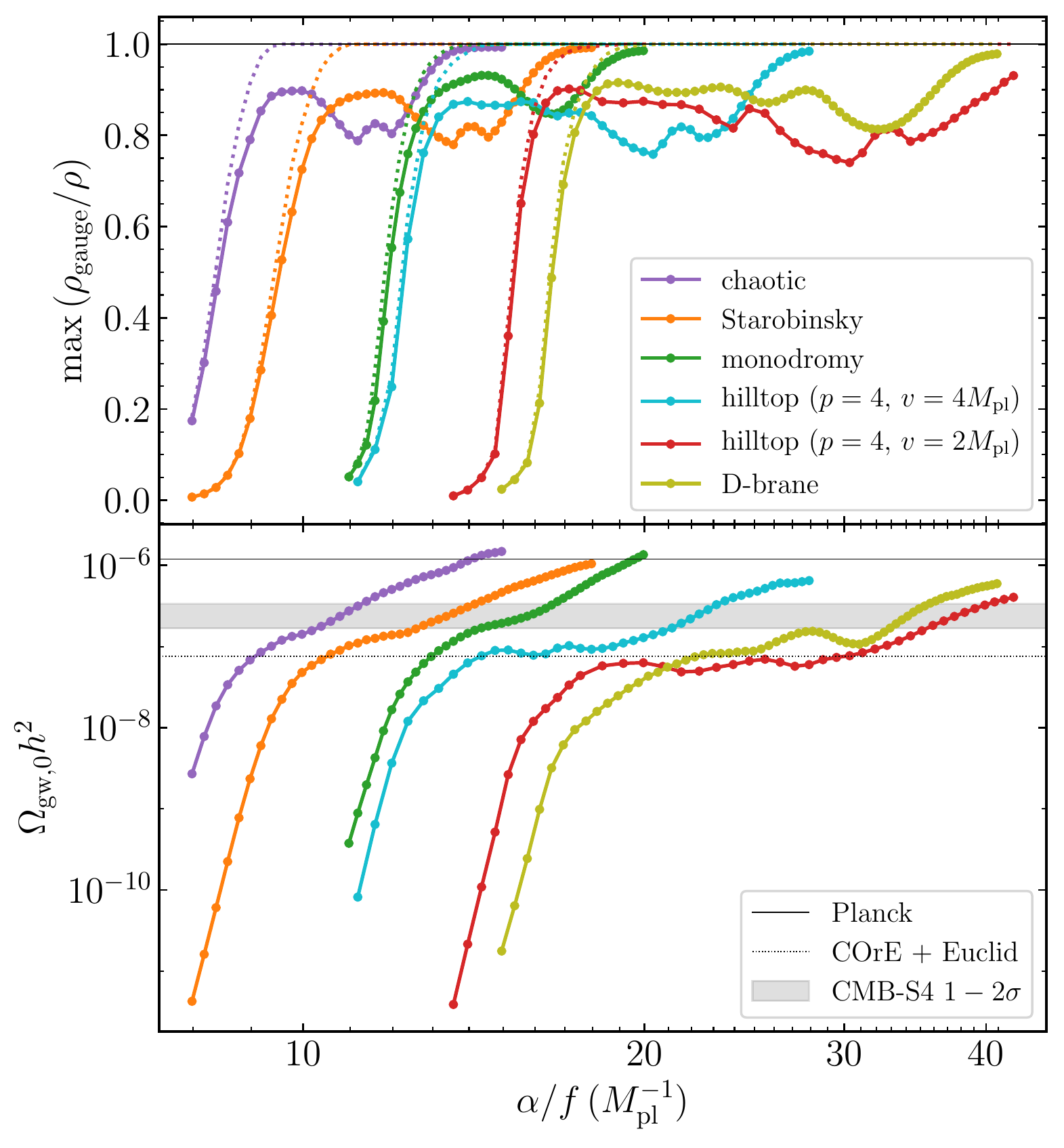}
    \caption{
        Preheating efficiency, quantified by the maximum $\rho_\mathrm{gauge} / \rho$ over the simulation (top panel), and the total fractional energy in gravitational waves today, $\Omega_{\mathrm{gw}, 0} h^2$ (bottom panel), as functions of axion-gauge coupling $\alpha / f$.
        Each color denotes a different inflationary potential, as labeled in the legend.
        The dashed lines (which follow the same color scheme) correspond to the maximum total energy in fluctuations, i.e., the maximum amount of energy in the simulation outside of the inflaton condensate.
        Lines indicating the current $\Delta N_\mathrm{eff}$ bounds on $\Omega_{\mathrm{gw}, 0} h^2$ from Planck and CMB-S4 from Ref.~\cite{Pagano:2015hma} are plotted in solid and dashed black, respectively, while the region between CMB-S4's $1 \sigma$ and $2 \sigma$ projections~\cite{Abazajian:2019eic} is shaded grey.
        Note that, to save clutter, we only plot a subset of the hilltop parameter points listed in \cref{tab:hilltop-model-fits}.
    }\label{fig:gw-money-vs-model}
\end{figure}

In \cref{fig:gw-money-vs-model}, we plot the efficiency of preheating and corresponding gravitational wave production over a range of axion--gauge-field couplings $\alpha / f$.
We first observe that the relationship between $\max \rho_\mathrm{gauge} / \rho$ and $\alpha / f$ follows the same general trend as presented in \cref{sec:dependence-inflationary-scale}.
That is, once $\alpha / f$ is large enough for preheating to be efficient, $\max \rho_\mathrm{gauge} / \rho$ remains roughly $90\%$ until backscattering effects become important, at which point we observe a dip in efficiency.
Finally, even larger couplings lead again to a regime of strong backreaction leading to slow tachyonic resonance, resulting in near-completely efficient preheating.
Turning to the lower panel of \cref{fig:gw-money-vs-model}, we see that i) depending on the coupling strength, preheating in all models can yield a net production of gravitational waves that would be probed by CMB-S4 measurements of $N_\mathrm{eff}$, and ii) models with $r \gtrsim 10^{-2}$ are already constrained by Planck data.
While all models we study here reach a regime of gravitational wave production detectable by future experiments such as CMB-S4, for those with $r \gtrsim 10^{-2}$ the entire regime of efficient preheating could be ruled out by a null detection of $\Delta N_\mathrm{eff}$.

\begin{figure}[t]
    \centering
    \includegraphics[width=\columnwidth]{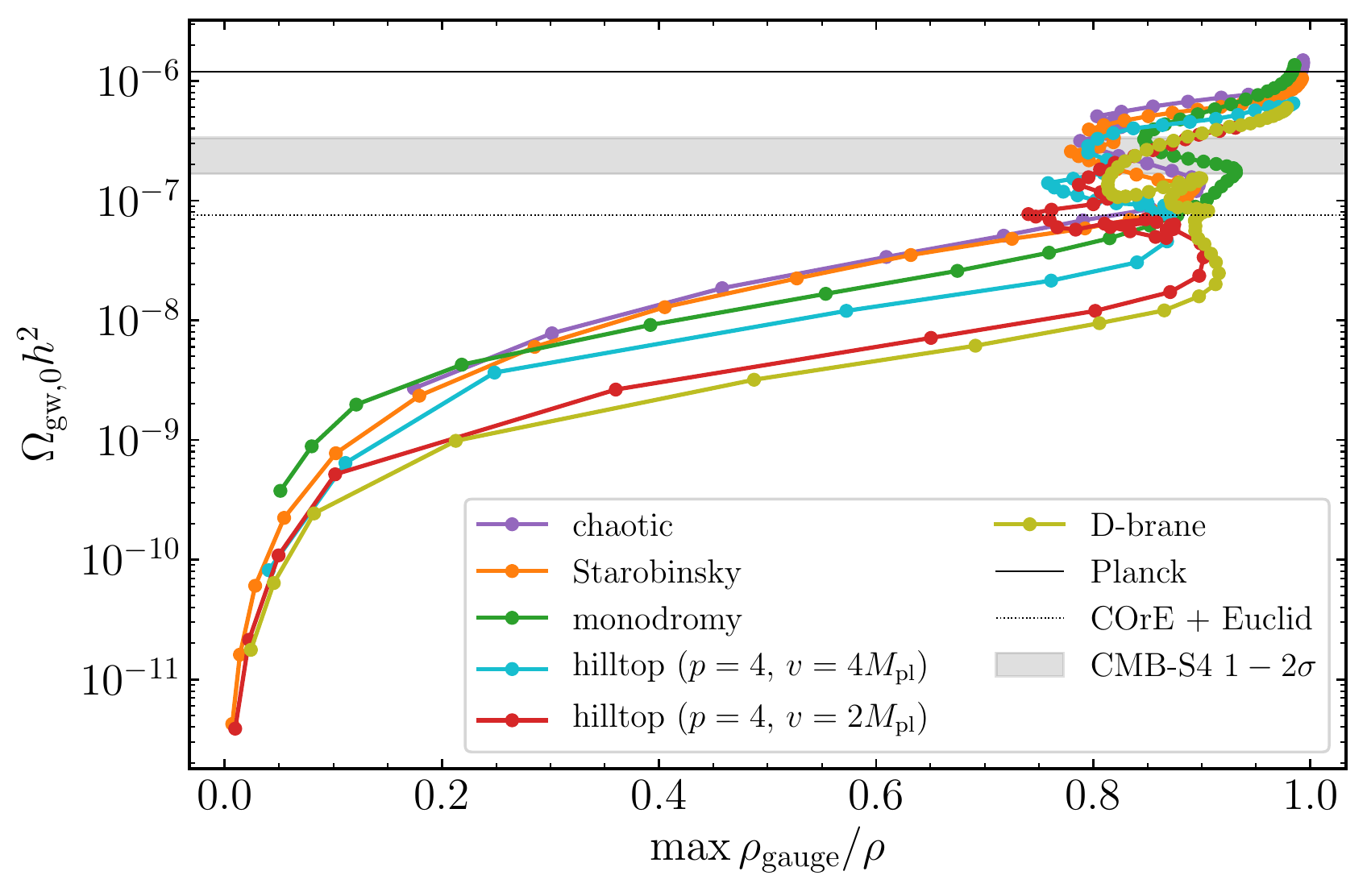}
    \caption{
        The total gravitational wave energy today versus the final preheating efficiency $\rho_\mathrm{gauge} / \rho$ for different inflationary potentials as indicated by the legend.
    }\label{fig:all-models-gw-total-vs-max-frac}
\end{figure}

In \cref{fig:all-models-gw-total-vs-max-frac} we display the energy density in gravitational waves today, $\Omega_{\mathrm{gw}, 0} h^2$, as a function of the efficiency of gauge preheating, which, as above, we quantify by the maximum fraction of energy in gauge field fluctuations during the simulation.
These results exhibit the sensitivity of $\Omega_{\mathrm{gw}, 0} h^2$ to the details of the potential.
That is, while the general relationship between gravitational wave production and preheating efficiency follows a similar trend, the overall scaling of $\Omega_{\mathrm{gw}, 0} h^2$ differs from model to model as observed in \cref{sec:dependence-inflationary-scale}.
Again, in general the larger coupling $\alpha / f$ required for complete preheating for a particular model, the less efficient the subsequent gravitational wave production.
Referring to \cref{tab:model-params}, low-scale models require larger coupling for efficient preheating and present correspondingly weaker detection prospects.
This observation, together with \cref{fig:gw-money-vs-model,fig:all-models-gw-total-vs-max-frac}, leads us to make the broad claim that for models of inflation with tensor-to-scalar ratios observable by CMB-S4 experiments, preheating into gauge fields could be simultaneously probed via the contribution of gravitational waves to $\Delta N_\mathrm{eff}$.

A positive detection of $r \gtrsim 10^{-3}$ together with $\Delta N_\mathrm{eff}$ by CMB-S4 experiments could provide evidence for a pseudoscalar inflaton (axion) reheating the Universe through preheating to gauge fields.
Alternatively, detection of $r$ with a measurement of $\Delta N_\mathrm{eff}$ consistent with zero would provide stringent bounds on the axion-gauge coupling $\alpha / f$---in particular the regime in which preheating is the sole mechanism by which the Universe was reheated would be ruled out.
In the former case, a precise prediction of the end of inflation (relative to the pivot scale used to parametrize CMB observables) may require the methods of \cref{sec:initial-conditions} to accurately model the backreaction of gauge fields onto the inflationary background toward the end of inflation.

\begin{figure}[t]
    \centering
    \includegraphics[width=\columnwidth]{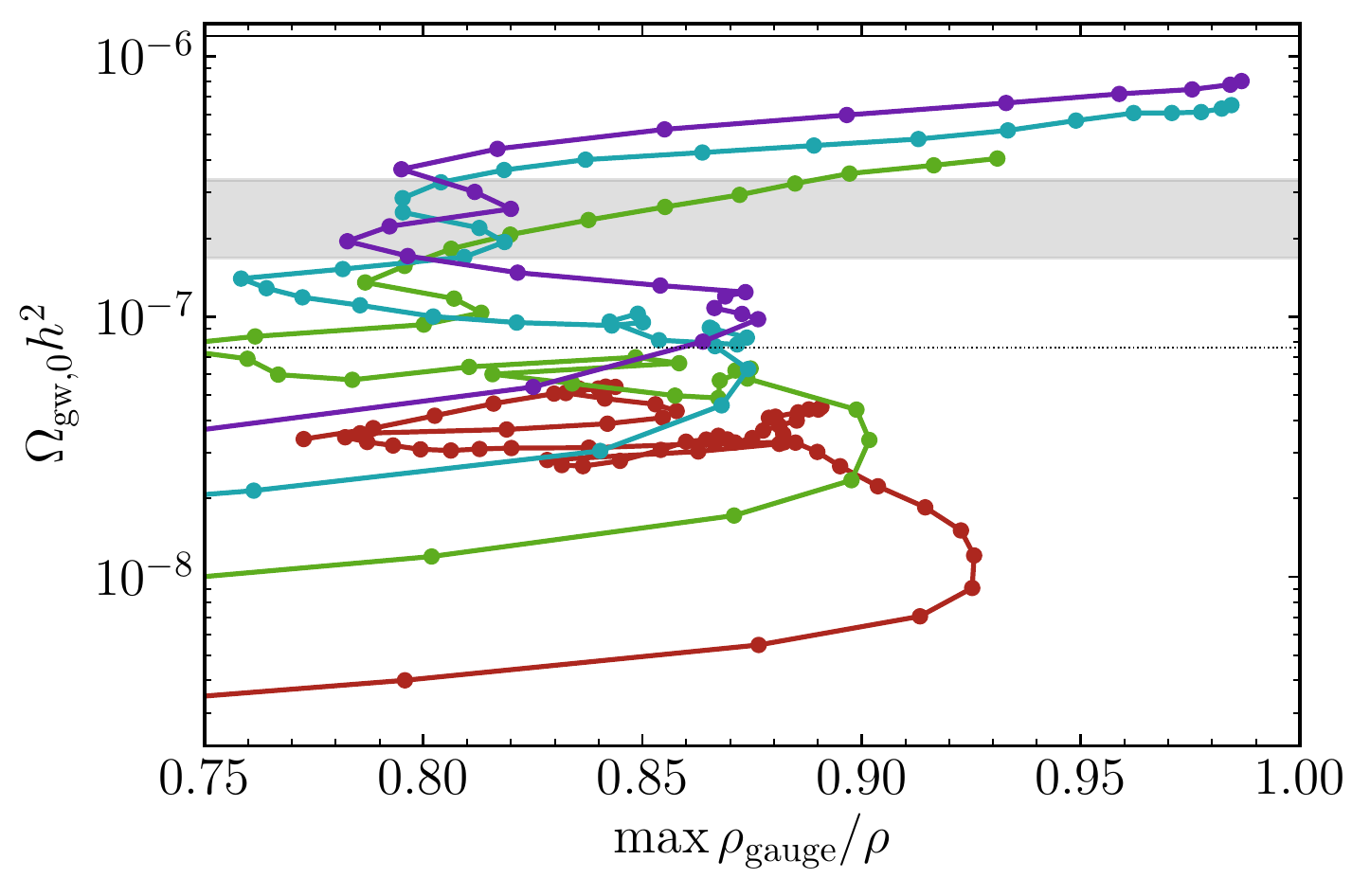}
    \caption{
        The total gravitational wave energy today versus the final preheating efficiency $\rho_\mathrm{gauge} / \rho$.
        Each color denotes the hilltop model with $p = 4$ and values of $v$ equal to $\Mpl$, $2 \Mpl$, $4 \Mpl$, and $8 \Mpl$ corresponding to colors red through purple.
    }\label{fig:gw-money-hilltop-vary-v}
\end{figure}

To evaluate the claim that the efficiency of gravitational wave production is correlated with $r$, we vary the free parameter $v$ of the hilltop model, which allows us to tune the flatness of the potential during inflation, and so both the energy scale of inflation and $r$.
In \cref{fig:gw-money-hilltop-vary-v} we observe the exact trend we noted in our survey of different potentials: while the relationship between $\max \rho_\mathrm{gauge} / \rho$ and $\Omega_{\mathrm{gw}, 0} h^2$ is clearly universal, as $v$ increases, gravitational wave production is more efficient (as is preheating, which requires increasingly lower $\alpha / f$ to be efficient).
Consulting \cref{tab:model-params}, increasing $v$ decreases $k_\ast / H_\ast$, which (referring to \cref{tab:hilltop-model-fits}) correlates to larger tensor-to-scalar ratios, $r$.
From \cref{fig:gw-money-hilltop-vary-v} we extrapolate that, regardless of the choice of $v$, at sufficiently large coupling the hilltop models will exhibit complete preheating (entering the regime of slow tachyonic resonance sustained by gauge-field backreaction).
However, the resulting gravitational wave production will be suppressed, reducing the constraining power of $\Delta N_\mathrm{eff}$ in this scenario.
We note, however, that existing constraints on the instability parameter $\xi$ during inflation (the strongest of which are due to primordial black hole production~\cite{Linde:2012bt, Bugaev:2013fya, Garcia-Bellido:2016dkw, Domcke:2017fix, Garcia-Bellido:2017aan, Cheng:2018yyr}) still provide significantly weaker constraints on $\alpha / f$ for small-$r$ (and so small-$\epsilon_H$) models, as $\partial \phi / \partial N \sim \sqrt{\epsilon_H}$.
At the same time, for larger values of $v$ there is a regime of detectable gravitational waves from preheating (while still $r \lesssim 10^{-3}$, below the target for CMB-S4).
As such, a detection of nonzero $\Delta N_\mathrm{eff}$ but not nonzero $r$ would be consistent with axion inflation in models with $r \lesssim 10^{-3}$.
Finally, we note that we have repeated this analysis for the D-brane model, finding the same relationship as for the hilltop model.


\section{Conclusions}\label{sec:Conclusions}

In this paper, we have studied the effect of the inflationary potential on the dynamics of gauge preheating and gravitational wave production after axion inflation.
Abelian gauge fields coupled via $\phi F \tilde{F}$ are amplified via a tachyonic resonance from the postinflationary axion condensate, which rapidly and efficiently reheats the Universe and copiously produces a stochastic background of gravitational waves.
For strong enough coupling, these signals, first computed in Ref.~\cite{Adshead:2018doq}, contribute so greatly to the radiation content of the Universe that Planck data already places bounds on the axion-gauge coupling $\alpha / f$, which CMB-S4 will improve upon significantly.
Here we have extended this study to consider a variety of inflationary models, finding that such a level of gravitational wave production typically occurs in scenarios where preheating completely reheats the Universe.

While the results are qualitatively similar across models, the efficiency of preheating (at a given coupling $\alpha / f$) varies depending on both the energy scale of inflation and the shape of the potential.
The scale of the potential (as well as the coupling parameter $\alpha / f$) controls the significance of backreaction of the gauge fields onto the axion's background dynamics.
At strong enough couplings, this gauge-field friction supports an extended phase of slow tachyonic resonance which persists until $100\%$ of the axion's energy has transferred to the gauge fields.
As such, lower-scale inflation models require larger coupling to enter into this regime, which suppresses the resulting gravitational wave amplitude by moving the source further inside the horizon.

The tensor-to-scalar ratio is sensitive to the flatness of the potential at CMB, measured by $\epsilon_H$, and (once the amplitude of scalar fluctuations is set to match CMB data) to the energy scale of inflation.
As a result, gravitational wave production is comparatively less efficient in models with smaller tensor-to-scalar ratios.
However, when preheating completely transitions the Universe to radiation domination, generally the level of gravitational wave production contributes to the effective number of radiationlike degrees of freedom $N_\mathrm{eff}$ at a level that could be detected (or ruled out) by next-generation CMB experiments.
Should CMB-S4 (which targets $r \gtrsim 10^{-3}$) detect nonzero $r$, a simultaneous detection of nonzero $\Delta N_\mathrm{eff}$ could be an indication that preheating occurred via strong coupling to gauge fields.
Conversely, the lack of observed gravitational waves via $\Delta N_\mathrm{eff}$ would place severe constraints on $\alpha / f$, in particular ruling out most of the parameter space in which preheating in this model is solely responsible for the transition to radiation domination.
Notably, for these models CMB-S4 could rule out the entire regime of couplings for which preheating alone reheated the Universe.

Planck data currently set $\alpha / f \lesssim 14 \, \Mpl^{-1}$ and $19.6 \, \Mpl^{-1}$ for the chaotic and monodromy potentials, respectively, which CMB-S4 will improve to $\alpha / f \lesssim 9 \, \Mpl^{-1}$ and $13 \, \Mpl^{-1}$.
\unskip\footnote{These constraints are often phrased in terms of the instability parameter $\xi$, defined in \cref{eqn:xi-definition}, evaluated when CMB modes left the horizon.
The Planck limits using our results are $\xi_\mathrm{CMB} \lesssim 0.9$ for both the chaotic and monodromy potentials; from CMB-S4 projections these are both $\sim 0.6$.
}
These results improve upon bounds from larger scales: non-Gaussianity limits $\alpha / f \lesssim 32.3 \Mpl^{-1}$ and $\alpha / f \lesssim 46.5 \, \Mpl^{-1}$, respectively~\cite{Barnaby:2010vf,Barnaby:2011vw}, while for primordial black hole (over) production these constraints are $\alpha / f \lesssim 21.9 \Mpl^{-1} - 24.9 \, \Mpl^{-1}$ and $\alpha / f \lesssim 35.9 \, \Mpl^{-1}$~\cite{Linde:2012bt, Bugaev:2013fya}.
Our results place similarly tight constraints for Starobinsky inflation (evident in \cref{fig:gw-money-vs-model}) and natural inflation (similar to those for chaotic inflation).

The couplings we constrain here do not exclude regimes that could be relevant for magnetogenesis~\cite{Adshead:2016iae} or gravitational leptogenesis~\cite{Adshead:2018doq}.
Further, non-Gaussianities can be generated during preheating~\cite{Enqvist:2004ey,Chambers:2007se,Chambers:2008gu}; however, there is no indication the scenario studied here would have a significant effect on observable scales.

The significant gravitational response (of the tensor part of the metric) to the preheating dynamics studied here motivates a similar investigation into the metric's scalar degrees of freedom (i.e., curvature perturbations) and the associated potential for primordial black hole formation.
Nonlinear gravitational effects could potentially alter our findings here, which may be studied in a framework similar to that employed by Ref.~\cite{Giblin:2019nuv}.
Additionally, we could investigate the extension of our results to more complicated gauge theories~\cite{Adshead:2017xll}.
Finally, we have ignored the effects of the backreaction of any charged matter on the gauge preheating process.
Charged particles are produced via the Schwinger effect, leading to a nonzero conductivity in the resulting plasma which could possibly damp gauge field production~\cite{Domcke:2018eki,Sobol:2019xls}.
We leave a detailed investigation of these questions for future work.


\acknowledgments
We thank Mustafa Amin and Jessie Shelton for useful discussions and Valerie Domcke for comments on our draft.
Z.J.W. thanks Andreas Kloeckner for generous support and advice on the development of \textsf{pystella}.
The work of P.A.\ was supported in part by NASA Astrophysics Theory Grant No. NNX17AG48G.
J.T.G.\ is supported by the National Science Foundation Grant No. PHY-1719652.
M.P. acknowledges the support of the Spanish MINECOs ``Centro de Excelencia Severo Ochoa'' Programme under Grant No. SEV-2016-059.
This project has received funding from the European Unions Horizon 2020 research and innovation programme under the Marie Sk\l{}odowska-Curie Grant Agreement No. 713366.
Z.J.W.\ is supported in part by the United States Department of Energy Computational Science Graduate Fellowship, provided under Grant No. DE-FG02-97ER25308.
The development of \textsf{pystella} made use of the Extreme Science and Engineering Discovery Environment (XSEDE)~\cite{xsede} through allocation TG-PHY180049, which is supported by National Science Foundation Grant No. ACI-1548562, and also made use of hardware purchased by the National Science Foundation, Kenyon College, and the Kenyon College Department of Physics.
This work made use of the Illinois Campus Cluster, a computing resource that is operated by the Illinois Campus Cluster Program (ICCP) in conjunction with the National Center for Supercomputing Applications (NCSA) and which is supported by funds from the University of Illinois at Urbana-Champaign.
P.A.\ acknowledges the hospitality of the Yukawa Institute for Theoretical Physics at Kyoto University, where some of this work was completed during the YITP-T-19-02 on ``Resonant instabilities in cosmology.''

\appendix


\section{Equations of motion and linear analysis}\label{App:eomsandlinalaysis}

In this appendix, we write down the equations of motion for the system and collect some well-known results about the linear evolution of the system during inflation and the backreaction of the gauge field on the inflaton.


\subsection{Gravitational sector}

The evolution of the background FLRW spacetime [\cref{eqn:flrw-metric}] is governed by the Einstein equations
\begin{align}
    \label{friedmann_1}
    \mathcal{H}(\tau)^2
    &= \frac{a{(\tau)}^2}{3 \Mpl^2} \rho(\tau) \\
   \label{friedmann_2}
    \mathcal{H}'(\tau)
    &= - \frac{a{(\tau)}^2 }{2 \Mpl^2} \left( 3 \rho(\tau) + p(\tau) \right),
\end{align}
where $\tau$ is the conformal time and $\mathcal{H} = a'/a$ is the conformal Hubble parameter related to the usual Hubble parameter via $\mathcal{H} = a H$.
For the action in \cref{eqn:action}, the spatially averaged energy density and pressure are
\begin{align}
    \label{rho}
    \rho(\tau)
    &\equiv \left\langle \frac{{\phi'}^2}{2 a^2} + \frac{(\partial_i \phi)^2}{2 a^2}+ V(\phi) + \frac{1}{2} \left( {\mathbf{E}}^2 + {\mathbf{B}}^2 \right) \right\rangle, \\
    \label{pressure}
    p(\tau)
    &\equiv \left\langle \frac{{\phi'}^2}{2 a^2} - \frac{(\partial_i \phi)^2}{6 a^2} - V(\phi) + \frac{1}{6} \left( {\mathbf{E}}^2 + {\mathbf{B}}^2 \right) \right\rangle,
\end{align}
where brackets $\langle \cdots \rangle$ denote a spatial average and we have defined the electric and magnetic fields
\begin{align}
    \label{eq:electric_magnetic}
    E_{i}
    &= \frac{1}{a^2} \left( {A_{i}}' - \partial_{i} A_0 \right), \quad
    B_{i}
    = \frac{1}{a^2} \epsilon_{i j k} \partial_{j} A_{k}.
\end{align}
Finally, our Fourier convention is set by
\begin{align}
    f(\mathbf{k})
    &= \int \ud^{3} x \, f(\mathbf{x}) e^{- i \mathbf{k} \cdot \mathbf{x}}, \\
    f(\mathbf{x})
    &= \int \frac{\ud^{3} k}{(2\pi)^3} \, f(\mathbf{k}) e^{i \mathbf{k} \cdot \mathbf{x}}.
\end{align}


\subsection{Gravitational wave dynamics}\label{sec:gw}

To study the production of gravitational waves, we consider an FLRW metric including tensor (but not scalar nor vector) perturbations,
\begin{align}
    \label{eq:spacetime_metric}
    \ud s^2
    &= a(\tau)^2 \left( - \ud \tau^2 + \left(\delta_{ij} + h_{ij}\right) \ud x^i \ud x^j \right),
\end{align}
where $\partial_i h_{ij} = h_{ii} = 0$ is the transverse-traceless perturbation of the spatial metric.
In this work we compute the dynamics of $h_{ij}$ passively---that is, we determine the amount of gravitational waves sourced by the axion and the gauge fields but we neglect their backreaction onto the system.

From the linearized Einstein equations we obtain the equation of motion for $h_{ij}$,
\begin{align}\label{eqn:GWinhomo}
	h''_{ij} - \partial_k \partial_k h_{ij} + 2 \mathcal{H} h'_{ij}
    - 2 \left( 2 \mathcal{H}' + \mathcal{H}^2 \right) h_{ij}
    &= \frac{2}{\Mpl^2} T_{ij}^\mathrm{TT}.
\end{align}
In this expression, $T_{ij}^\mathrm{TT}$ is the transverse-traceless component of the stress-energy tensor,
\begin{align}
    \label{TT-project}
	T_{ij}^\mathrm{TT} = \left(P_{il} P_{jm} - \frac{1}{2} P_{ij} P_{lm} \right) T_{lm},
\end{align}
where the transverse-traceless projector is
\begin{align}\label{projector}
	P_{ij}
    &= \delta_{ij} - \frac{k_i k_j}{k^2}.
\end{align}
The background Einstein equations show that $2 \left( 2 \mathcal{H}' + \mathcal{H}^2 \right)$ is proportional to the background pressure $p(\tau)$.
Since usually all the modes in our simulations are subhorizon, this term in \cref{eqn:GWinhomo} induces a negligible amount of dispersion and so we neglect it.

Under the approximations discussed after \cref{eq:spacetime_metric}, the stress-energy tensor is expressed as
\begin{align}
    \label{gw-source}
    \begin{split}
        T_{ij}
        &= \partial_i \phi \partial_j \phi + F_{i\alpha} F_{j\beta} \bar{g}^{\alpha\beta} \\
        &\hphantom{={}}
        - \bar{g}_{ij} \left[ \frac{1}{2} \partial_\mu \phi \partial^\mu \phi + V(\phi) + \frac{1}{4} F_{\mu\nu} F^{\mu\nu} \right],
    \end{split}
\end{align}
where $\bar{g}_{\mu\nu}$ denotes the unperturbed FLRW metric.
Moreover, since the terms proportional to $\bar{g}_{ij}$ are pure trace, they do not contribute to \cref{TT-project} (they are projected out by $P_{ij}$).
As a consequence, only the first two terms in \cref{gw-source} are responsible for gravitational wave generation.

The stress tensor of gravitational waves is~\cite{Misner:1974qy}
\begin{align}
	T_{\mu\nu}^\mathrm{gw} = \frac{\Mpl^2}{4} \left\langle h_{ij,\mu} h_ {ij,\nu} \right\rangle,
\end{align}
with the sum over $i$ and $j$ implied.
The corresponding fractional energy density in gravitational waves as
\begin{align}\label{GWdensitySpec}
	\Omega_\mathrm{gw}(k)
	&\equiv \frac{1}{\rho} \frac{\ud \rho_\mathrm{gw}}{\ud \ln k} \\
	&= \frac{1}{24\pi^2 L^3} \frac{k^3}{\mathcal{H}^2} \sum_{i, j} \left\vert h_{ij}^\prime(k, \tau)) \right\vert^2.
\end{align}
The spectrum redshifted to today is related to the spectrum at emission by the transfer function~\cite{Easther:2006vd}
\begin{align}\label{gw-transfer-function}
	\Omega_\mathrm{gw,0}(f) h^2
	&= \Omega_{\mathrm{gw},e}(f) \left(\frac{g_0}{g_\ast}\right)^{1/3} \Omega_{\mathrm{r},0} h^2
\end{align}
at frequencies
\begin{align}\label{frequencyToday}
	f \approx 2.7 \times 10^{10} \frac{k_\mathrm{phys}}{ \sqrt{\Mpl H}} \, \mathrm{Hz},
\end{align}
where $k_\mathrm{phys}$ is the physical wave number and $H$ is the Hubble parameter evaluated at the time when the spectrum is being computed.
Above, $g_0 / g_\ast$ is the ratio of thermal degrees of freedom today to matter-radiation equality.
Here we assume that the Universe was radiation dominated from the moment of emission until matter-radiation equality.


\subsection{Field equations}\label{sec:field-equations}

The Euler-Lagrange equations for the action \cref{eqn:action} yield the dynamics of the gauge fields,
\begin{align}\label{field-tensor-eom}
    \partial_{\mu} \left( \sqrt{-g} F^{\mu \nu} + \sqrt{-g} \frac{\alpha}{f} \tilde{F}^{\mu \nu} \right) = 0.
\end{align}
In terms of the gauge potentials, the $\nu = i$ equations are the dynamical equations of motion
\begin{align}
    \label{eqn:gauge-eom}
    \begin{split}
        0 &=
        {A_i}'' - \partial_i A_0' - \partial_j \partial_j A_i + \partial_i \partial_j A_j \\
        &\hphantom{={}}
        - \frac{\alpha}{f} \epsilon_{i k l} \phi' \partial_k A_{l} + \frac{\alpha}{f} \epsilon_{i k l} \partial_k \phi \left( A_l' - \partial_l A_0 \right),
    \end{split}
\end{align}
while the $\nu = 0$ equation is the Gauss constraint
\begin{align}
    \label{gauss-law}
    \partial_i {A_i}' - \partial_j \partial_j A_0
    &= - \frac{\alpha}{f} \epsilon_{i j k} \partial_{k} \phi \partial_{i} A_{j}.
\end{align}
The four components of the gauge field are not physical: the theory is invariant under gauge transformations where
\begin{align}
    A_{\mu} \to \tilde{A}_\mu = A_{\mu} - \partial_\mu \beta,
\end{align}
where $\beta$ is an arbitrary function.
This freedom allows us to eliminate one degree of freedom by fixing a gauge.
Together with Gauss's law, the gauge sector represents only two physical, dynamical degrees of freedom.

Likewise, the Euler-Lagrange equation for the inflaton provides its equation of motion,
\begin{align}
    \label{eq:phi-eom}
	\phi'' - \partial_i \partial_i \phi + 2 \mathcal{H} \phi' + a^2 \frac{\ud V}{\ud \phi}
    &= - a^2 \frac{\alpha}{4 f} F_{\mu \nu} \tilde{F}^{\mu \nu}.
\end{align}
Note that $F_{\mu \nu} \tilde{F}^{\mu \nu} / 4 = a^4 \mathbf{E} \cdot \mathbf{B}$.


\subsection{Linear theory}

The system described by \cref{field-tensor-eom,eq:phi-eom} is difficult to solve analytically due to the nonlinear interactions between the axion and gauge field.
However, provided the coupling $\alpha/f$ is not too large, a linear treatment provides some insight into the dynamics of this system during inflation and the early phase of preheating.

We begin by linearizing the system of equations, expanding the axion about its homogeneous background $\phi = \bar{\phi}(\tau) + \delta\phi$ and treating the gauge field as a first-order perturbation.
We choose the temporal gauge, $A_0 = 0$.
Note that, at linear order this gauge is equivalent to Coulomb gauge $\partial_i A_i = 0$ via the gauge constraint; the gauge field may be taken to be purely transverse.
Expanding into Fourier modes
\begin{align}
    A_i(\tau,\mathbf{x})
    &= \sum_{\lambda = \pm} \int \frac{\ud^3 k}{(2\pi)^3}
        A_k^{\lambda}(\tau) \varepsilon^\lambda(\mathbf{k}) e^{- i \mathbf{k} \cdot \mathbf{x}}
\end{align}
where the polarization vectors $\varepsilon^\pm(\mathbf{k})$ form an orthogonal basis of polarizations transverse to the momentum $\mathbf{k}$,
\begin{align}
	\epsilon_{ijk} k_j \varepsilon^\pm_k(\mathbf{k}) = \mp i k \varepsilon^\pm_i(\mathbf{k}),
\end{align}
and satisfy the relations
\begin{align}\label{pol-vec-orthonormal}
	\varepsilon^\lambda_i(\mathbf{k}) \varepsilon^{\lambda'}_i(\mathbf{k})^\ast
    &= \delta^{\lambda \lambda'}, \\
	k_i \varepsilon^\pm_i(\mathbf{k})
    &= 0, \\
    \varepsilon^\pm_i(-\mathbf{k})
    &= \varepsilon^\pm_i(\mathbf{k})^\ast, \\
    \varepsilon^\pm_i(\mathbf{k})^\ast
    &= \varepsilon^\mp_i(\mathbf{k}).
\end{align}
At linear order, the equation of motion, \cref{eqn:gauge-eom}, reduces to
\begin{align}
    \label{eqn:linear-eom-gauge-polarization-appendix}
    {A_\pm}''(\mathbf{k}) + k \left( k \mp 2 \mathcal{H} \xi \right) A_\pm(\mathbf{k}) = 0,
\end{align}
where we have defined the instability parameter
\begin{align}
	\label{eqn:xi-definition}
	\xi \equiv \frac{\alpha}{2 f} \frac{\phi'}{ \mathcal{H} }.
\end{align}
\Cref{eqn:linear-eom-gauge-polarization-appendix} shows that the interaction between the inflaton and the gauge fields induces a tachyonic instability (for modes with $k/\mathcal{H} < 2 \vert \xi \vert$) for one of the two helicity states ($A_+$ if $\phi' > 0$, or conversely $A_-$ if $\phi' < 0$).
The modes within this band experience (for values of $\xi \gtrsim 2$) exponential enhancement slightly before horizon crossing and then approach a constant value on superhorizon scales.
In the limit of a nearly constant $\xi$, the solutions to this equation are Whittaker functions (see, e.g.,~\cite{Adshead:2015pva}).
Assuming $\phi' > 0$, near horizon crossing ($k / a H \sim 1$) these solutions are well approximated by
\begin{align}
    \label{eq:A_approx}
    A_+ \simeq \frac{1}{\sqrt{2k}} \left( \frac{k}{2 \mathcal{H} \xi }\right)^{1/4} e^{ \pi \xi - 2 \sqrt{2 \xi k/\mathcal{H} }}.
\end{align}
These results evince the importance of the instability parameter $\xi$ which controls the exponential enhancement of the gauge field.

Computing the backreaction of the gauge field onto the background dynamics requires the integrals
\begin{align}
    \label{E2_integral}
    \frac{1}{2} \left\langle \mathbf{E}^{2} \right\rangle
    &=\frac{1}{4 \pi^{2} a^{4}} \sum_{\lambda=\pm} \int \ud k \, k^{2} \left\vert A_{\lambda}^{\prime}(k)\right\vert^{2} \\
    \label{B2_integral}
    \frac{1}{2} \left\langle \mathbf{B}^{2} \right\rangle
    &=\frac{1}{4 \pi^{2} a^{4}} \sum_{\lambda=\pm} \int \ud k \, k^{4}\left\vert A_{\lambda}(k)\right\vert^{2}, \\
    \label{EdotB_integral}
    \left\langle \mathbf{E} \cdot \mathbf{B} \right\rangle
    &= - \frac{1}{4 \pi^{2} a^{4}} \sum_{\lambda=\pm} \lambda \int \ud k \, k^{3} \partial_\tau \left\vert A_{\lambda}(k) \right\vert^{2},
\end{align}
where $A_{\lambda}(k)$ denotes the helicity modes of the gauge potentials.
(Note that \cref{E2_integral,B2_integral,EdotB_integral} assume the temporal gauge.)
For $\xi \gtrsim 4$~\cite{Jimenez:2017cdr}, fairly accurate approximations of these quantities are given by~\cite{Anber:2009ua}
\begin{align}
    \label{EdotB_analytic}
    \left\langle \mathbf{E} \cdot \mathbf{B} \right\rangle
    &\simeq 2.4 \times 10^{-4} \, \frac{\mathcal{H}^4}{a^4 \xi^4} e^{2 \pi \xi}, \\
    \label{E2B2_analytic}
    \frac{1}{2} \left\langle \mathbf{E}^2 + \mathbf{B}^{2} \right\rangle
    &\simeq 1.4 \times 10^{-4} \, \frac{\mathcal{H}^4}{a^4 \xi^3} e^{2 \pi \xi}.
\end{align}
Substituting \cref{EdotB_analytic} into \cref{eq:phi-eom} demonstrates that gauge fields exert an additional friction term for the inflaton (see \cref{friction_comparison}).
Since $\xi$ is proportional to the inflaton background's velocity, it is expected to grow during slow-roll inflation.
As such, backreaction may significantly alter the evolution of the inflaton toward the end of inflation, even before preheating.
In \cref{sec:initial-conditions} we present our procedure to accurately account for these effects during inflation (as used to set initial conditions for the lattice simulations).


\section{Initial conditions}\label{sec:initial-conditions}

As implemented in Refs.~\cite{Adshead:2015pva,Adshead:2016iae,Adshead:2018doq}, in order to capture the tachyonic enhancement of (one polarization of) the gauge fields during inflation, we numerically integrate the linearized equations of motion for the fluctuations of the gauge-field helicity modes, \cref{eqn:linear-eom-gauge-polarization-appendix}, during inflation.
These fluctuations are integrated alongside the background dynamics, i.e., \cref{friedmann_2}, and the homogeneous part of the inflaton's equation of motion, \cref{eq:phi-eom}.
To set initial conditions for the subsequent lattice simulation, we evaluate the background quantities and obtain power spectra from the integrated gauge field modes between one and two $e$-folds before $\epsilon_H = 1$, marking the end of inflation, as described in \cref{sec:numerical-methods}.
By beginning the lattice simulation sufficiently early before the end of inflation we capture nonlinear effects that become important, and thus only use the solutions to the linearized equations while they remain valid.

To account for the gauge fields' backreaction onto the background dynamics during inflation (as described in \cref{sec:field-equations}), Refs.~\cite{Adshead:2015pva,Adshead:2016iae,Adshead:2018doq} use the approximations \cref{EdotB_analytic,E2B2_analytic}.
The accuracy of these expressions is sensitive to two main assumptions.
First, the solution of \cref{eqn:linear-eom-gauge-polarization-appendix} is well approximated by the Whittaker function only in the limit that $\xi$ is constant, and $\mathcal{H} = -1/\tau$ (in near--de Sitter space).
Second, \cref{EdotB_analytic,E2B2_analytic} themselves are approximations to the integrals \cref{E2_integral,B2_integral,EdotB_integral} over \cref{eq:A_approx}, accurate for $\xi \gtrsim 4$~\cite{Jimenez:2017cdr}.
When $\xi$ is nearly constant but small, the approximate integrals \cref{EdotB_analytic,E2B2_analytic} are inaccurate, but the Whittaker solution is still valid.
However, if $\xi$ is not approximately constant, then the Whittaker solution itself is inaccurate, meaning the \textit{integrands} of \cref{E2_integral,B2_integral,EdotB_integral} themselves are inaccurate, regardless of the size of $\xi$.

At low couplings (which roughly correspond to those which do not achieve complete preheating), the gauge fields, while amplified, have a negligible effect on the background evolution during inflation.
Thus, while $\xi$ is not large enough for \cref{EdotB_analytic,E2B2_analytic} to be valid, their effect is small enough that this error is unimportant.
At larger couplings for which preheating is complete, the gauge fields have a more substantial impact on the background evolution during the final $e$-folds of inflation, but these effects can in principle be captured by linearized calculations (without further approximations).
However, in this regime analytic approximations are in fact never simultaneously valid: during slow roll when the Whittaker solution is approximately valid, $\abs{\xi} < 4$, meaning the approximation to the analytic integral is not trustworthy~\cite{Jimenez:2017cdr}.
In fact, by the time $\xi$ is large enough to validate the approximation to the integral, the inflaton is no longer slowly rolling, so the Whittaker solution is no longer valid.
Further, the backreaction of the gauge fields onto the background dynamics is no longer negligible (during the final $e$-folds of inflation), meaning the background evolution using these approximations may be inaccurate.
At even stronger couplings, nonlinear dynamics have an important impact on the background evolution during the final $e$-folds of inflation.
The friction exerted by the gauge fields on the inflaton background (beyond what is captured by a linear treatment) can even postpone the end of inflation.
In this regime we might not trust any calculation using only the linear equations of motion toward the end of inflation, so we leave a detailed study of such strong coupling to future work.

Here we improve upon this procedure by computing the gauge-field integrals \cref{E2_integral,B2_integral,EdotB_integral} via numerical quadrature of the numerically evolved gauge fields.
As chosen for the analytic results \cref{EdotB_analytic,E2B2_analytic}, we take as a UV cutoff for these integrals $k = 2 a H \xi$, the upper end of the tachyonic instability band; we find that, in the regime where the gauge fields' contribution to the background evolution is non-negligible, the integration is insensitive to slight variation about this cutoff.
With this method, the evolution of the background in tandem with the linearized fluctuations agrees very well with the background dynamics of the subsequent lattice calculation (for the overlapping final one to two $e$-folds of inflation).

In practice, it is prohibitively expensive to include the full dynamic range of fluctuations which are amplified---the band of tachyonic instability $k \lesssim 2 a H \xi$ is proportional to the comoving horizon $\mathcal{H} = a H$, which increases exponentially during inflation.
To address this, we only include modes which experience amplification during the last $\sim 20$ $e$-folds of inflation, well before backreaction has any impact on the background dynamics of inflation (for the values of $\alpha / f$ we consider).
We sample this range of modes (spanning 8 to 10 orders of magnitude) logarithmically, yielding good results from quadrature of \cref{E2_integral,B2_integral,EdotB_integral} over $\ln k$.
A final complication is that fluctuations deep inside the horizon oscillate extremely rapidly compared to the Hubble rate.
However, this regime ($k \gg 2 a H \xi$) is precisely the regime where the Wentzel-Kramers-Brillouin (WKB) approximation
\begin{align}
    A_\pm(k, \tau) \approx \frac{1}{\sqrt{2 \omega_\pm(k, \tau)}} e^{i \omega_\pm(k, \tau) \tau}
\end{align}
with $\omega_\pm(k, \tau)^2 = k^2 \mp 2 a H \xi k$ is extremely accurate.
Thus, we only numerically integrate a given mode $k$ when $k \leq 50 \times 2 a H \xi$, using the WKB solution until this point.
Using an adaptive ordinary differential equation integrator (in practice, \textsf{SciPy}'s DOP853 routine~\cite{scipy,10.5555/153158}) ensures that we take as large of time steps as possible while maintaining a prescribed relative accuracy (typically to one part in $10^{11}$).
A similar strategy was employed in Ref.~\cite{Cheng:2015oqa} to study this model's dynamics during both inflation and preheating; however, the validity of such a method applied to study preheating is unclear.

\begin{figure*}[t]
    \centering
    \includegraphics[width=\textwidth]{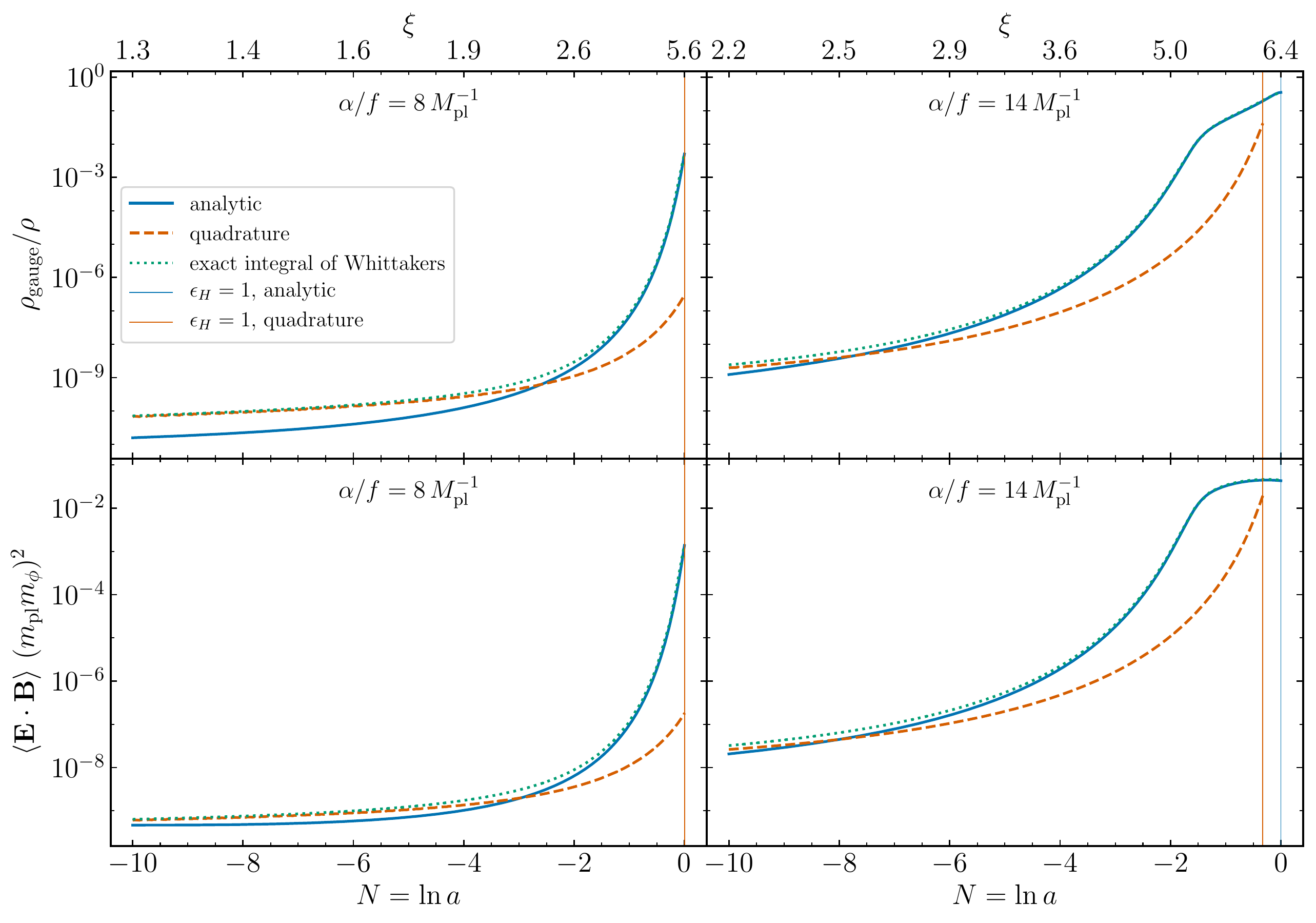}
    \caption{
        Comparison of the analytic approximations (solid blue lines) to $\langle \mathbf{E}^2 + \mathbf{B}^2 \rangle$ (top panels) and $\langle \mathbf{E} \cdot \mathbf{B} \rangle$ (bottom panels) (i.e., \cref{EdotB_analytic,E2B2_analytic}) to those obtained via direct numerical quadrature of gauge field fluctuations evolved with their linearized equations of motion (dashed orange lines).
        Both background evolutions are initialized identically, with $N = 0$ corresponding to the time inflation ends (i.e., $\epsilon_H = 1$) in the evolution using the analytic approximations.
        The vertical lines indicate the point at which inflation ends according to both evolutions.
        We additionally plot the result of direct numerical quadrature of \cref{E2_integral,B2_integral,EdotB_integral} using the Whittaker-function solutions to evaluate the accuracy of \cref{EdotB_analytic,E2B2_analytic} as approximations to this exact integral.
    }\label{gauge-exp-vals-compare}
\end{figure*}

In \cref{gauge-exp-vals-compare} we investigate the difference between the analytic approximations [\cref{EdotB_analytic,E2B2_analytic}] and the numerical computations described above, fixing a chaotic inflaton potential.
We consider couplings where the background evolutions agree well ($\alpha / f = 8 \, \Mpl^{-1}$) and disagree ($\alpha / f = 14 \, \Mpl^{-1}$).
We additionally compare to the numerical quadrature of \cref{E2_integral,B2_integral,EdotB_integral} using the actual Whittaker-function solutions, which demonstrates agreement with the approximations \cref{EdotB_analytic,E2B2_analytic} once $\abs{\xi} \gtrsim 3 - 4$.
Earlier in inflation, when exact slow roll (and so the Whittaker-function solution) is a good approximation, we observe agreement between the exact integrals of the Whittaker function solutions and the integrals over numerically evolved gauge field perturbations.
However, in neither case do the analytic approximations agree with the results of numerical evolution and quadrature at any time.

While these errors have no significant effect on the inflationary background deep in inflation, \cref{gauge-exp-vals-compare} indicates that the analytic approximations greatly overestimate $\langle \mathbf{E}^2 + \mathbf{B}^2 \rangle$ and $\langle \mathbf{E} \cdot \mathbf{B} \rangle$ during the final two $e$-folds of inflation, regardless of the coupling to the axion.
In the strong-coupling case (where gauge-field backreaction is significant), the analytic result's overestimation of the friction exerted by the gauge fields onto the axion background leads to inflation ending artificially late, as indicated by the vertical lines (marking when $\epsilon_H = 1)$ in \cref{gauge-exp-vals-compare}.
In this case, with the analytic-based evolution we would end up initializing the lattice simulation with background values roughly half an $e$-fold later than intended (depending on the coupling).
Note that $\alpha / f = 8 \, \Mpl^{-1}$ is roughly the largest coupling for which the background evolutions visibly agree.


\section{Post-preheating dynamics}\label{sec:post-preheating-dynamics}

In this appendix we study the dynamics after preheating to evaluate whether the Universe remains radiation dominated.
In particular, we seek to quantify the amount by which the fractional energy density in gravitational waves is suppressed due to any deviation from radiation domination between preheating and the end of reheating.

As discussed in \cref{sec:dependence-inflationary-scale}, at the end of all of our simulations a small amount of the total energy remains in axion fluctuations [$\mathcal{O}(1\%)$ for the cases where preheating is most efficient].
While these are typically relativistic at this point, because the axion is rather massive these fluctuations become nonrelativistic within just a few $e$-folds and gravitate as pressureless matter.
Depending on their lifetime, these massive axions can come to dominate the energy density of the Universe, leading to a period of matter domination.
Even if the Universe does not become matter dominated, the Universe could still depart significantly from radiation domination.
The resulting effect on the gravitational wave transfer function depends on (i) how much energy remains in the axion, $\Omega_\phi$, and (ii) its decay rate into radiation, $\Gamma$.

Noting that $\rho_\mathrm{GW} \propto a^{-4}$ for the modes generated inside the horizon during preheating, the fractional density in gravitational waves scales as
\begin{align}\label{eqn:gw-redshift-eos}
    \Omega_{\mathrm{gw}}
    \equiv \frac{\rho_\mathrm{gw}}{\rho}
    \propto \frac{1}{(a / a_0)^4 (H / H_0)^2}
    \propto \left( \frac{a}{a_0} \right)^{3 w - 1},
\end{align}
where a subscript $0$ denotes some reference time (which for our purposes is the end of preheating) and $w \equiv p / \rho$ is the time-dependent equation of state.
Thus, if the Universe remains radiation dominated ($w = 1/3$), then $\Omega_{\mathrm{gw}}$ remains constant (as it redshifts at the same rate as the rest of the Universe).
On the other hand, if the Universe is dominated by matter ($w = 0$) or by a mixture of matter and radiation ($0 < w < 1/3$), then $\Omega_{\mathrm{gw}}$ decreases with the scale factor.
Thus, any suppression of gravitational waves after preheating depends on both the (evolution of the) equation of state $w$ and the duration for which $w < 1/3$.

In the standard reheating scenario (without a preheating phase), the equation of state of the Universe is that of matter due to the inflaton condensate's oscillation about the minimum of its potential.\footnote{More precisely, the equation of state of a coherently oscillating scalar field time averages to zero.}
Matter domination persists until the decay of the inflaton into relativistic species (in our case, bosons) becomes efficient, which occurs when the inflaton's decay rate $\Gamma$ becomes comparable to present the Hubble scale $H$.
Accounting for preheating affects this description in two significant ways: the equation of state is not (initially) $w = 0$ and the coupled sector is already (highly) occupied.
The former would delay the possible onset of matter domination before reheating completes, while the latter significantly alters the inflaton decay rate due to Bose enhancement.
Accounting for the nontrivial phase space distribution of the gauge fields $f_A(p)$ in our models enhances the decay rate by a factor $\sim 1 + 2 f_A(m_\phi / 2)$~\cite{Adshead:2016xxj}.

For simplicity, we study the dynamics after preheating with the standard Boltzmann equations describing three-body decay of inflaton particles at rest into two relativistic daughter particles~\cite{Kolb:1990vq},
\begin{align}\label{eqn:boltzmann-phi}
    \frac{\ud \rho_\phi}{\ud t} + 3 H \rho_\phi
    &= - \Gamma \rho_\phi \\
    \label{eqn:boltzmann-gamma}
    \frac{\ud \rho_\gamma}{\ud t} + 4 H \rho_\gamma
    &= \Gamma \rho_\phi.
\end{align}
Rather than fixing the decay rate $\Gamma$ and the initial (i.e., post-preheating) fractional energy in the axion $\Omega_\phi$ according to our model (and the final state of our simulations), we instead study a wide range of parameter space and use order-of-magnitude estimates to determine whether this period significantly affects our results.

\begin{figure}[t]
    \centering
    \includegraphics[width=\columnwidth]{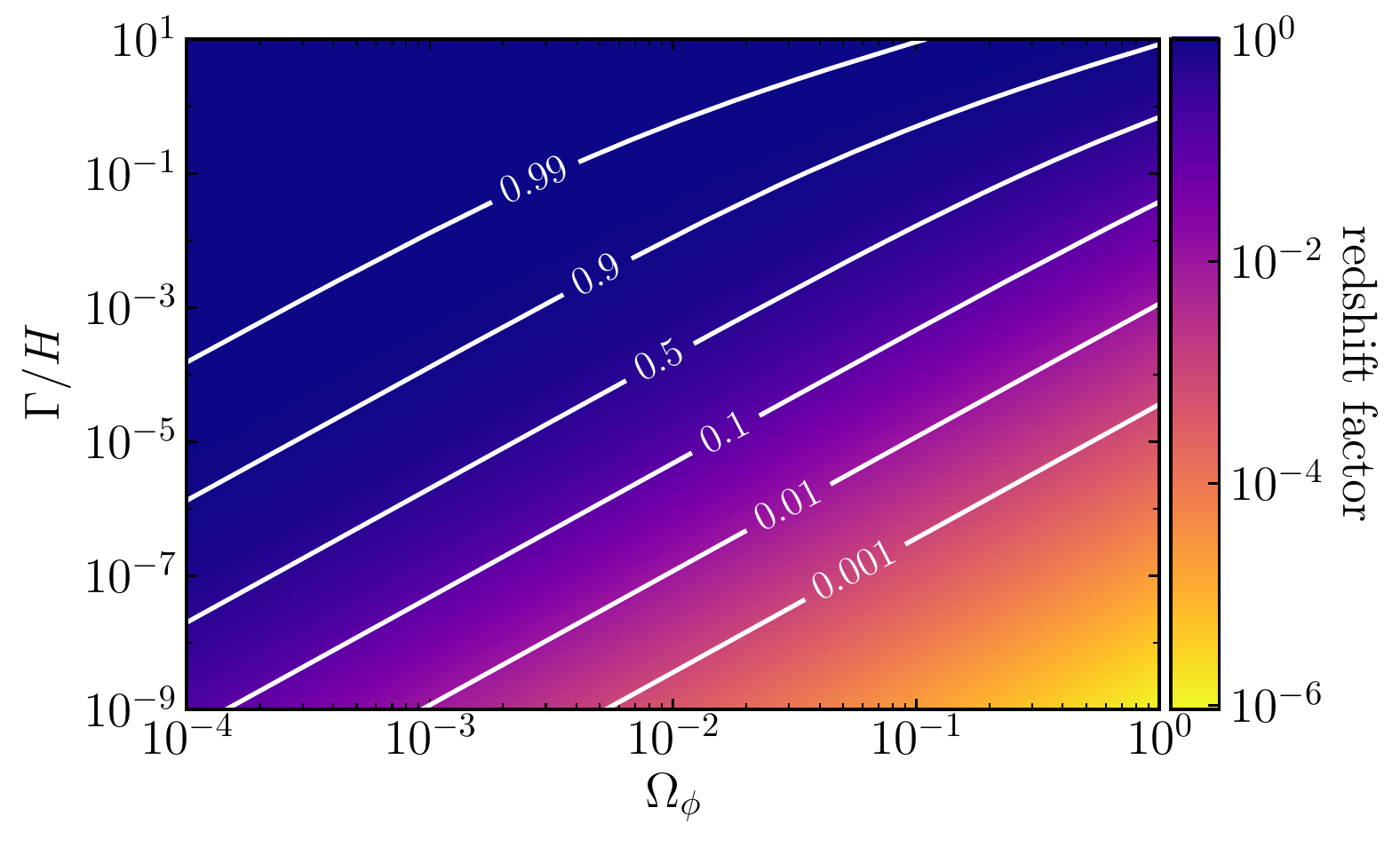}
    \caption{
        The amount by which $\Omega_{\mathrm{gw}}$ would decrease between preheating and the time the Universe fully reheats, as a function of the fraction of the Universe's energy remaining in the axion after preheating, $\Omega_\phi$, and the effective decay rate $\Gamma$ of the axion into a relativistic species (relative to the Hubble rate $H$ at the end of preheating).
    }\label{fig:redshift-factor-vs-Gamma-Omega-phi}
\end{figure}

In \cref{fig:redshift-factor-vs-Gamma-Omega-phi} we numerically integrate \cref{eqn:boltzmann-phi,eqn:boltzmann-gamma} over a number of decades in $\Gamma$ and $\Omega_\phi$, plotting the amount by which $\Omega_{\mathrm{gw}}$ would reduce due to $w$ being less than $1/3$ during this epoch (using \cref{eqn:gw-redshift-eos}).
In this figure, the vertical slice at $\Omega_\phi = 1$ corresponds to reheating without any preheating phase.
In our model, the zero-temperature decay rate of the axion into gauge fields is~\cite{Tanabashi:2018oca}
\begin{align}
    \frac{\Gamma_0}{H} = \frac{1}{H} \frac{\alpha^2 m_\phi^3}{64 \pi f^2} \sim 10^{-9}.
\end{align}
To estimate $f_A(m_\phi / 2)$ we extract the occupation number of the gauge fields from the simulations, which varies between $\sim 10^{8}$ and $10^{12}$, in line with results from tachyonic resonance in scalar-field preheating~\cite{Dufaux:2006ee}.
Thus, the relevant portion of \cref{fig:redshift-factor-vs-Gamma-Omega-phi} is $10^{-2} \lesssim \Omega_\phi \lesssim 10^{-1}$ and $\Gamma \gtrsim 10^{-1}$, for which the fraction energy density in gravitational waves would drop by no more than $\sim 10\%$, and likely by less than $1\%$.
Naturally, the regime for which $\Omega_{\mathrm{gw}, 0} h^2$ exceeds the projected CMB-S4 bounds on $\Delta N_\mathrm{eff}$ corresponds to both the smallest $\Omega_\phi$ (approaching $1\%$) and the largest occupation numbers.
Therefore, our results are most robust for the coupling regimes which would be probed or excluded by CMB-S4.

In principle, the true dynamics depend on the time dependence of the entire phase-space distribution $f(p)$, as well as other possible thermal effects due to sectors coupled to the daughter particles.
We have further approximated that the axion particles are at rest (while at the end of the simulations all would have nonzero momentum).
However, \cref{fig:redshift-factor-vs-Gamma-Omega-phi} makes clear that we are far from the regime where $\Omega_{\mathrm{gw}}$ is suppressed by any appreciable amount: the effective decay rate could drop by $\sim 2$ to $8$ orders of magnitude before $\Omega_{\mathrm{gw}}$ would redshift to half of its original value.
As such, we conclude that the Universe would remain radiation dominated after preheating, and so our reported $\Omega_{\mathrm{gw}, 0} h^2$ are robust.


\bibliography{axion-gw}

\end{document}